\documentclass[a4paper,11pt]{article}
\pdfoutput=1
\usepackage{amssymb}
\usepackage{a4wide}
\usepackage{psfrag}
\usepackage{amsmath,euscript,array,mathrsfs}
\usepackage{epstopdf}
\usepackage{longtable}
\usepackage{enumerate}
\usepackage{color}
\usepackage{hyperref}

\numberwithin{equation}{section}
\usepackage[numbers]{natbib}
\usepackage{ifthen}
\newcommand{\beq}{\begin{equation}}
\newcommand{\eeq}{\end{equation}}
\newcommand{\beqs}{\begin{eqnarray}}
\newcommand{\eeqs}{\end{eqnarray}}

\renewcommand{\L}{{\cal L}}

\def\hbar{\hspace{0pt}\raisebox{1pt}{$-$} \hspace{-7pt} h}

\def\r{\rho}

\newcommand{\be}{\begin{equation}}
\newcommand{\ee}{\end{equation}}
\newcommand{\bea}{\begin{eqnarray}}
\newcommand{\eea}{\end{eqnarray}}
\newcommand{\nn}{\nonumber}

\def\tr{{\rm tr}}

\def\lbldef#1#2{\expandafter\gdef\csname #1\endcsname {#2}}

\def\href#1#2{#2}

\newcommand{\ber}{\begin{eqnarray}}
\newcommand{\eer}{\end{eqnarray}}

\newcommand{\beqar}{\begin{eqnarray}}

\newcommand{\eeqar}{\end{eqnarray}}
\newcommand{\tht}{\thteta}

\newcommand{\dsl}
  {\kern.06em\hbox{\raise.15ex\hbox{$/$}\kern-.56em\hbox{$\partial$}}}

\newcommand{\D}{{\cal{D}}}

\newcommand{\N}{{\cal N}}

\newcommand{\eeqarr}{\end{eqnarray}}
\newcommand{\ZZ}{{\rm \kern 0.275em Z \kern -0.92em Z}\;}

  \def\tr{{\hbox{\rm Tr}}}

\def\CC{{\mathchoice
{\rm C\mkern-8mu\vrule height1.45ex depth-.05ex
width.05em\mkern9mu\kern-.05em}
{\rm C\mkern-8mu\vrule height1.45ex depth-.05ex
width.05em\mkern9mu\kern-.05em}
{\rm C\mkern-8mu\vrule height1ex depth-.07ex
width.035em\mkern9mu\kern-.035em}
{\rm C\mkern-8mu\vrule height.65ex depth-.1ex
width.025em\mkern8mu\kern-.025em}}}

\def\RR{{\rm I\kern-1.6pt {\rm R}}}

\def\ZZ{{\rm Z}\kern-3.8pt {\rm Z} \kern2pt}
\def\IB{\relax{\rm I\kern-.18em B}}
\def\ID{\relax{\rm I\kern-.18em D}}
\def\II{\relax{\rm I\kern-.18em I}}
\def\IP{\relax{\rm I\kern-.18em P}}

\newcommand{\bear}{\begin{eqnarray}}
\newcommand{\eear}{\end{eqnarray}}

\def\to{\rightarrow}

\def\tr{{\rm Tr}}
\def\to{\rightarrow}

\def\a{\alpha}

\def\k{\kappa}                    
\def\l{\lambda}

  \def\w{\omega}
  \def\th{\theta}                  
\def\r{\rho}                                     

\def\6{\partial}

\def\bea{\begin{eqnarray}}
\def\eea{\end{eqnarray}}

\def\beqx{\begin{displaymath}}
\def\eeqx{\end{displaymath}}

\newcommand{\bmat}{\left(\begin{array}}
\newcommand{\emat}{\end{array}\right)}

\def\a{\alpha}

\def\k{\kappa}
\def\l{\lambda}

    \def\th{\theta}
\def\r{\rho}

\def\D{\Delta}

\def\L{\Lambda}

\def\bo{{\raise-.3ex\hbox{\large$\Box$}}}               
\def\face{{\raise.2ex\hbox{$\displaystyle \bigodot$}\mskip-2.2mu \llap {$\ddot
        \smile$}}}                                   
\def\>{\rangle}                                      
\def\<{\langle}                                      

\def\wt#1{\widetilde{#1}}                            
\def\leftrightarrowfill{$\mathsurround=0pt \mathord\leftarrow \mkern-6mu
        \cleaders\hbox{$\mkern-2mu \mathord- \mkern-2mu$}\hfill
        \mkern-6mu \mathord\rightarrow$}        
\def\dvec#1{\vbox{\ialign{##\crcr
        \leftrightarrowfill\crcr\noalign{\kern-1pt\nointerlineskip}
        $\hfil\displaystyle{#1}\hfil$\crcr}}}           
\def\tr{{\rm tr \,}}                                    

\def\-{\hphantom{-}}

\definecolor{highlight}{rgb}{1,1,0.8}

\def\eqrefs#1#2{(\ref{#1}--\ref{#2})}

\def\eqrefc#1#2{(\ref{#1},~\ref{#2})}

\newcommand{\sech}{\mathop{\mathrm{sech}}\nolimits}

\newcommand{\rMN}{\rho_{h_1}}
\newcommand{\rSUSY}{\rho_\text{SUSY}}

\newcommand{\vph}{\varphi}

\newcommand{\Nc}{N_\text{c}}
\newcommand{\hh}{{\hat{h}}}

\renewcommand{\tht}{{\tilde\theta}}
\newcommand{\pht}{{\tilde\varphi}}
\renewcommand{\wt}{{\tilde\omega}}

\newcommand{\Koo}{K_{00}}
\newcommand{\Kto}{K_{20}}

\newcommand{\I}{{\mathcal{I}}}

\begin{document}

\baselineskip=15.5pt
\pagestyle{plain}
\setcounter{page}{1}
\begin{titlepage}

\vspace*{6mm}

\begin{center} 
\Large \bf The Structure of the Non-SUSY Baryonic Branch of Klebanov-Strassler
\end{center}

\vskip 10mm
\begin{center}
Stephen~Bennett\footnote{pystephen@swansea.ac.uk} and 
Daniel~Schofield\footnote{pyschofield@swansea.ac.uk}
 \vskip 4mm

{\it Department of Physics, Swansea University\\
 Singleton Park, Swansea SA2 8PP, United Kingdom.}
\vskip 5mm

\vspace{0.2in}
\end{center}
\begin{center}
{\bf Abstract}
\vskip0.5truecm
\begin{minipage}{0.9\textwidth}
We study the two-dimensional space of supergravity solutions corresponding to non-supersymmetric deformations of the baryonic branch of Klebanov-Strassler. By combining analytical methods with a numerical survey of the parameter space, we find that this solution space includes as limits the softly-broken $\N=1$ solutions of Gubser et al. and those of Dymarsky and Kuperstein. We also identify a one-dimensional family of solutions corresponding to a natural non-supersymmetric generalisation of Klebanov-Strassler, and one corresponding to the limit in which supersymmetry is completely absent, even in the far UV. For almost all of the parameter space we find indications that much of the structure of the supersymmetric baryonic branch survives.


\end{minipage}
\end{center}
\vskip1truecm
\vspace{0.1in}
\end{titlepage}
\setcounter{footnote}{0}

\renewcommand{\theequation}{{\rm\thesection.\arabic{equation}}}
%



\section{Introduction}\label{intro}
One of the primary aims of the study of gauge/gravity duality is to find a dual description of realistic field theories such as QCD. This requires generalising the original AdS/CFT correspondence \cite{Maldacena:1997re} to cases with less supersymmetry. For example, the Klebanov-Strassler (KS) \cite{Klebanov:2000hb} and Chamseddine-Volkov/Maldacena-N\'u\~nez (CVMN) \cite{Maldacena:2000yy,Chamseddine:1997nm} backgrounds constitute exact globally regular solutions which are dual to $\N=1$ gauge theories.

The presence of some remaining supersymmetry played a critical role in these successes, both in simplifying the search for solutions and in guaranteeing their stability. Despite this, considerable progress has been made with respect to the problem of finding dual descriptions which completely lack supersymmetry. One natural way that this can be achieved is by finding solutions in which a black hole is present, corresponding to a gauge theory at finite temperature \cite{Witten:1998zw}. See for example \cite{Buchel:2000ch,Buchel:2001gw,Gubser:2001ri,Gubser2001}.

Alternatively, one can consider field theories in which supersymmetry is softly broken by the insertion of relevant operators into the Lagrangian. By using as a starting point theories for which the duality is well understood, it is then possible to find dual gravity theories which are deformations of the SUSY case, as was achieved, for example, in \cite{Gubser2001,Evans:2002mc,Aharony:2002vp,Apreda:2003gc,Borokhov:2002fm,Babington:2002ci,
Babington:2002qt,Apreda:2003gs,Dymarsky:2011ve}. Specifically, the deformed background will match the original one asymptotically in the UV. The fact that the deformed backgounds share many of the features, such as symmetries, of the SUSY solutions means that the problem of finding solutions is considerably simplified.

In particular this approach was used in \cite{Bennett:2011va} to obtain non-SUSY solutions by deforming backgrounds on the baryonic branch of KS \cite{Butti:2004pk}. Although the SUSY system reduces to a single second-order differential equation, in the non-SUSY case it is necessary to solve the full equations of motion, consisting of six coupled second-order equations. It was practical to solve for asymptotic expansions essentially because of the similarities to the SUSY system --- the expansions have the same general form. By combining the expansions with numerical solutions it was possible to calculate several quantities in the dual field theory, and this confirmed that the behaviour was very similar to that in the SUSY case. 

In this paper we establish a more complete understanding of the space of solutions to which the solutions of \cite{Bennett:2011va} belong. This can be achieved in part as a result of the fact noted above, that the non-SUSY solutions share much of the structure of the SUSY baryonic branch. By consideration of the asymptotic expansions we find a two-dimensional parameter space which includes several previously studied solutions. In addition to the SUSY baryonic branch (and its limits, CVMN and KS itself), we also find the non-SUSY solutions of \cite{Gubser2001} and \cite{Dymarsky:2011ve} as limiting cases. By combining the structure described in \cite{Gubser2001} with that of the SUSY baryonic branch, it is possible to describe a generic non-SUSY solution in terms of transitions between regions in which SUSY and non-SUSY effects dominate. 

Additionally, we find some interesting special cases, one corresponding to a natural non-SUSY generalisation of KS itself, and the other to solutions in which SUSY is no longer \emph{softly} broken and the UV does not match the SUSY case asymptotically.

We begin in section \ref{SUSYsection} by reviewing relevant aspects of the SUSY baryonic branch and CVMN solutions. In section \ref{breakingSUSYsection} we turn to the non-SUSY solutions. First we review the solutions of \cite{Gubser2001}, obtained as a deformation of the CVMN background, before moving on to the main solutions of interest --- the generalisation of the baryonic branch obtained in \cite{Bennett:2011va}. Section \ref{2dSS} contains the main results of this paper. We first discuss the behaviour of generic solutions, and then concentrate on various special cases and limits. Finally we include some remarks on aspects of the dual field theory in section \ref{fieldTheory}.


\section{The SUSY system}\label{SUSYsection}
\subsection{Overview}\label{SUSYoverview}

Here we present two field theories, which although on the face of it appear different, are in fact connected via `higgsing' (as discussed in \cite{Maldacena:2009mw,Elander:2011mh}). The two theories are firstly that found when $\Nc$ D5-branes are wrapped on the 2-cycle of the resolved conifold (`theory~A'), and secondly the baryonic branch of the Klebanov-Strassler quantum field theory (`theory~B').

Theory~A is given by performing a special twisted compactification (to four dimensions), of six dimensional $SU(\Nc)$ supersymmetric Yang-Mills with 16 supercharges, preserving only 4 of them.  It was studied in \cite{Maldacena:2000yy,Andrews:2005cv,Andrews:2006aw} and has a field content (in the four dimensional language) consisting of a massless vector multiplet alongside a `Kaluza-Klein' tower of massive chiral and vector multiplets. The form of the Lagrangian, the weakly coupled mass spectrum and degeneracies of the theory are written in \cite{Andrews:2005cv,Andrews:2006aw}. The local and global symmetries are
\begin{align}
SU(\Nc)\times SU(2)_L\times SU(2)_R\times U(1)_R,
\end{align}
where the R-symmetry is anomalous, breaking $U(1)_R\to \mathbb{Z}_{2\Nc}$.

Theory~B is a quiver with gauge group $SU(n+\Nc)\times SU(n)$ and bifundamental matter multiplets $A_i, B_\alpha$ with $i,\a=1,2$. The global symmetries are (where again, the R-symmetry is anomalous)
\begin{align}
	SU(2)_L\times SU(2)_R \times U(1)_B \times U(1)_R .						\label{globalsymm}
\end{align}
These bifundamentals transform under the local and global symmetries as
\begin{align}
	A_i	=	\left( n+\Nc,\ 					\bar{n},\ 2,\ 1,\  1,\ \tfrac12 \right)	,	\qquad
	B_\a=	\left(\bar{n}+\bar\Nc,\	n			 ,\ 1,\ 2,\ -1,\ \tfrac12	\right).
\end{align}

There is a superpotential which can be written as $ W=\frac{1}{\mu} \epsilon_{ij} \epsilon_{\a\beta} \,\tr[A_i B_\a A_j B_\beta].$ The field theory is taken to be close to a strongly coupled fixed point at high energies and it can be shown that the anomalous dimension should be $\gamma_{A,B}\sim -\frac12$. This field theory is known to be dual to the Klebanov-Strassler (KS) background \cite{Klebanov:2000hb} and its generalization to the baryonic branch \cite{Butti:2004pk}.

The connection between theories A and B is via `higgsing' as mentioned above. If we give a (classical) baryonic vacuum expectation value to the fields $(A_i, B_\a)$ and then expand around it, we find that the degeneracies and field content of \cite{Andrews:2005cv,Andrews:2006aw} are recovered.

In terms of the Type IIB string backgrounds dual to each of the field theories, this weakly coupled field theory connection is manifest as a U-duality \cite{Maldacena:2009mw} (and was further studied in \cite{Elander:2011mh,Gaillard:2010qg,Minasian:2009rn,Halmagyi:2010st,Caceres:2011zn}). The first background (dual to theory~A) can be presented using the vielbeins
\begin{align}
E^{x_i}	 	&=	e^{\frac{\Phi}{4}}	dx_i					,\qquad
E^{\r}		 = e^{\frac{\Phi}{4}+k}d\r						,&
E^{\th}		&= e^{\frac{\Phi}{4}+h}d\th  					,\qquad
E^{\vph}	 = e^{\frac{\Phi}{4}+h}\sin\th d\vph 	,\nn\\ 
E^1			 	&= \frac12 e^{\frac{\Phi}{4}+g}(\wt_1 + a d\th)						,&
E^2				&=	\frac12	e^{\frac{\Phi}{4}+g}(\wt_2 - a \sin\th d\vph)	,\nn\\
E^3			 	&= \frac12	e^{\frac{\Phi}{4}+k}(\wt_3 + \cos\theta d\vph)
\label{vielbeinUnrot}
\end{align}

where we have used the following $SU(2)$ left-invariant 1-forms
\begin{align}
	\wt_1	&=	\cos\psi d\tht + \sin\psi\sin\tht d\pht,		\qquad\qquad \wt_2 = -\sin\psi d\tht + \cos\psi\sin\tht d \pht	\nn\\
	\wt_3	&=	d\psi + \cos \tht d\pht.
\label{LI1F}
\end{align}

This means we can write the background and the Ramond-Ramond 3-form compactly as
\begin{align}
	ds_E^2 &=	\sum_{i} (E^i)^2	,		\nn\\
	F_3		 &= e^{-\frac34 \Phi}	\Bigl[	f_1 E^{123}		+ f_2 E^{\th\vph 3}
																	+ f_3(E^{\th23}	+ E^{\vph13})
																	+ f_4(E^{\r1\th}+ E^{\r\vph2})
													   \Bigr],
\label{fieldsUnrot}
\end{align}
where we have defined
\begin{align}
E^{ijk\dots l}	&=	E^i \wedge E^j	\wedge E^k	\wedge \dots \wedge E^l ,	\nn\\
f_1							&=	-2 \Nc e^{-k-2g}																		, &
f_2							&=	\frac{\Nc}{2}e^{-k-2h}(a^2 -2 a b +1 )							,	\nn\\
f_3							&=	\Nc e^{-k-h-g}(a-b)																	, &
f_4							&=	\frac{\Nc}{2}e^{-k-h-g}b' 													.
\label{fDefs}
\end{align}
In this setup, the dilaton is a function $\Phi(\r)$ of the radial coordinate only, and we set $\a'g_s=1$. Then the background is written in terms of six functions, $\{ g,h,k, \Phi, a,b \}$, which all depend on the radial coordinate $\r$ only. It is possible to solve the SUSY system using a set of BPS equations that can be derived for the above ansatz.

The family of solutions we will present in section \ref{SUSYsols} correspond to a dual field theory deformed by the insertion of an eight-dimensional operator in the Lagrangian which couples the field theory to gravity. This calls for a completion in the context of the field theory which is achieved on the supergravity side with a U-duality \cite{Maldacena:2009mw}.  We will refer to this procedure as the `rotation'. It amounts to a solution generating technique which yields the `rotated' background, in which the vielbeins are
\begin{align}
e^{x_i}	 &=	e^{\frac{\Phi}{4}}	\hh^{-\frac14}dx_i					,\qquad
e^{\r}		= e^{\frac{\Phi}{4}+k}\hh^{\frac14} d\r						,&
e^{\th}	 &= e^{\frac{\Phi}{4}+h}\hh^{\frac14} d\th  				,\qquad
e^{\vph}	= e^{\frac{\Phi}{4}+h} \hh^{\frac14} \sin\th d\vph ,\nn\\ 
e^1			 &= \frac12 e^{\frac{\Phi}{4}+g} \hh^{\frac14} (\wt_1 + a d\th)					,&
e^2			 &=	\frac12	e^{\frac{\Phi}{4}+g} \hh^{\frac14} (\wt_2 - a \sin\th d\vph)	,\nn\\
e^3			 &= \frac12	e^{\frac{\Phi}{4}+k} \hh^{\frac14} (\wt_3 + \cos\theta d\vph).	
\label{vielbeinRot}
\end{align}
The `rotation' leaves the RR 3-form invariant\footnote{The factor of $\hh^{-3/4}$ in \eqref{fieldsRot} relative to in \eqref{fieldsUnrot} simply cancels the factors contained in the new vielbeins \eqref{vielbeinRot}. } but turns on some new fluxes. The new metric, RR and NS fields are then
\begin{align}
	ds_E^2	&=	\sum_{i} (e^i)^2	,		\nn\\
	F_3			&= \frac{e^{-\frac34 \Phi}}{\hh^{3/4}}
							\Bigl[	f_1 e^{123}		+ f_2 e^{\th\vph 3}
																	+ f_3(e^{\th23}	+ e^{\vph13})
																	+ f_4(e^{\r1\th}+ e^{\r\vph2})
					    \Bigr],							\nn\\
	H_3			&=	-\k \frac{e^{\frac54 \Phi}}{\hh^{3/4}}
							\Bigl[	- f_1 e^{\th\vph\r} - f_2 e^{\r12}
											- f_3 ( e^{\th2\r} + e^{\vph1\r} )
											+ f_4 ( e^{1\th3} + e^{\vph23} )
							\Bigr],							\nn\\
	C_4			&=	-\k \frac{e^{2\Phi}}{\hh} dt \wedge dx_1 \wedge dx_2 \wedge dx_3,		\nn\\
	F_5			&=	\k e^{-\frac54 \Phi - k}\hh^{\frac34}	\partial_\r \left( \frac{e^{2\Phi}}{\hh} \right)
							\left[	e^{\th\vph123} - e^{t x_1 x_2 x_3 \r}	\right].
\label{fieldsRot}
\end{align}
In the above equations we have a new factor defined as
\begin{align}
	\hat{h}=1-\k^2 e^{2\Phi}.				\label{rotationhh}
\end{align}
We choose the constant $\k$ to be such that the dual QFT will decouple from gravity (corresponding to careful removal of the eight-dimensional operator). The choice that allows this is $\k= e^{-\Phi_\infty}$, where $\Phi_\infty$ is the asymptotic value of the dilaton for large $\r$. This requirement restricts us to those solutions in which the dilaton is bounded at large distances. The rationale behind this choice is discussed in more detail in \cite{Elander:2011mh,Caceres:2011zn}.

\subsection{The SUSY solutions}\label{SUSYsols}

The background described in \eqrefs{vielbeinUnrot}{fDefs} results in a system of non-linear, coupled, first-order BPS equations (which are derived in the appendix of \cite{Casero:2006pt}). These can be repackaged using a certain change of basis functions \cite{Casero:2007jj,HoyosBadajoz:2008fw,Nunez:2010sf} into a much simpler form where the equations decouple: We rewrite the background functions $\{g,h,k, a,b\}$ in terms of five new functions $\{P,Q,Y,\tau,\sigma\}$ according to
\begin{align}
	4e^{2h}&= \frac{P^2-Q^2}{P\cosh\tau - Q}	,		&
	e^{2g} &=	P\cosh\tau - Q									,		&
	e^{2k} &=	4Y															,		\nn\\
	a			 &=	\frac{\sinh\tau}{P\cosh\tau - Q},		&
	\Nc b	 &=	\sigma													.
\end{align}
Then most of the BPS equations can be reduced to algebraic relations between the functions, leaving a single decoupled second-order equation for $P$ (referred to in the literature as the `master equation'): 
\begin{align}
	P'' + P' \left[		\frac{P'+Q'}{P-Q} + \frac{P'-Q'}{P+Q} - 4\coth(2\r - 2\r_o)		\right]		=	0,		\label{masterEq}
\end{align}
with
\begin{align}
	Q = (Q_o + \Nc)\coth(2\r-2\r_o) + \Nc\left[ 2\r \coth(2\r-2\r_o) - 1 \right],		\label{QEq}
\end{align}
where $Q_o$ and $\r_o$ are two integration constants. Each solution to the master equation \eqref{masterEq} generically provides us with two backgrounds, related by the U-duality or rotation described in section \ref{SUSYoverview}.\footnote{Or, more generally, a family of backgrounds parametrised by $\kappa$ in \eqref{rotationhh}.} We will be most interested in the rotated solutions, which correspond to the baryonic branch. However, much of what follows will be concerned simply with the behaviour of the background functions, and so will apply equally to the unrotated case (corresponding to theory A in section \ref{SUSYoverview}). Additionally, we will at times deal with solutions in which the dilaton grows without bound in the UV. Then, as discussed above, we can see from \eqref{rotationhh} that we cannot apply the rotation procedure without the warp factor $\hh$ vanishing.\footnote{This does not necessarily mean that we cannot consider these solutions as belonging to the rotated family. The issue is in fact slightly more subtle, and we will return to this point in section \ref{h1Dep}.}

The master equation \eqref{masterEq} describes all solutions compatible with the ansatz \eqrefs{vielbeinUnrot}{fDefs}. However, we will restrict our attention to globally regular solutions. In this case we find the solutions have an IR (for $\r\to0$) of the form
\begin{align}
	e^{2g}		&=	\frac{h_1}{2} 		+ \frac{4h_1}{15}\left(3 - \frac{5\Nc}{h_1} - \frac{2\Nc^2}{h_1^2}	\right)\r^2		+ O(\r^4),	\nn\\
	e^{2h}		&=	\frac{h_1}{2}\r^2 - \frac{4h_1}{45}\left(6 - \frac{15\Nc}{h_1} + \frac{16\Nc^2}{h_1^2}	\right)\r^4	+ O(\r^6),	\nn\\
	e^{2k}		&=	\frac{h_1}{2} 		+ \frac{2h_1}{5}\left(1 - \frac{4\Nc^2}{h_1^2}		\right)\r^2											+ O(\r^4),	\nn\\
	e^{\Phi-\phi_0}
						&=	1	+	\frac{16\Nc^2}{9h_1^2}\r^2																																			+ O(\r^4),	\nn\\
	a					&=	1 - \left( 2 - \frac{8\Nc}{3h_1} \right)\r^2																												+ O(\r^4),	\qquad\qquad
	b					 =	\frac{2\r}{\sinh 2\r}																																												 ,	\label{SUSYBBir}
\end{align}
where the exact expression for $b$ holds for all $\r$. Aside from the ability to shift the dilaton, encoded in $\phi_0$, we therefore have a family of solutions parametrised by $h_1$. The second integration constant we expect from the second-order equation \eqref{masterEq} has been fixed to ensure regularity. The same requirement also leads us to fix the values of the integration constants appearing in \eqrefs{masterEq}{QEq} as $Q_o=-\Nc$ and $\r_o=0$.

Turning to the UV, we find that for $\r\to\infty$
\begin{align}
	e^{2g}						&=	c_+ e^{\frac43 \r}		+	\Nc (1-2\r)
																							+	\frac{\Nc^2}{c_+}\left( \frac{13}{4} - 4\r + 4\r^2  \right)e^{-\frac43 \r}
																							+O\bigl( e^{-\frac83 \r} \bigr)	,		\nn\\
	e^{2h}						&=	\frac{c_+}{4} e^{\frac43 \r}		-	\frac{\Nc}{4} (1-2\r)
																												+	\frac{\Nc^2}{c_+}\left( \frac{13}{16} - \r + \r^2  \right)e^{-\frac43 \r}
																												+O\bigl( e^{-\frac83 \r} \bigr)	,		\nn\\
	e^{2g}						&=	\frac{2c_+}{3} e^{\frac43 \r}
																							-	\frac{\Nc^2}{3c_+}\left( \frac{25}{2} - 20\r - 8\r^2  \right)e^{-\frac43 \r}
																							+O\bigl( e^{-\frac83 \r} \bigr)	,		\qquad
	b									 =	\frac{2\r}{\sinh 2\r}																							\label{SUSYBBuv} \\ 
	e^{4(\Phi-\Phi_\infty)}									
										&=	1 + \frac{3\Nc^2}{4c_+^2} \left( 1 - 8\r  \right) e^{-\frac83\r}		+		O\bigl( e^{-\frac{16}{3}\r} \bigr),		\;\;
	a									 =	2 e^{-2\r} -\frac{2\Nc}{c_+} (1 - 8\r) e^{-\frac{10}{3}\r} +		O\bigl( e^{-\frac{14}{3}\r} \bigr),	\nn	
\end{align}
with an additional parameter $c_-$ appearing at the next order, giving two non-trivial parameters. Of course, we require a smooth solution joining the two expansions \eqrefs{SUSYBBir}{SUSYBBuv}, and this can be seen to be the case numerically. However, there is then only one non-trivial independent parameter; given a value for one of $\{h_1,c_+,c_-\}$, the requirement that the interpolating solution matches both the IR and UV expansions is sufficient to determine the values of the other two. This can be seen numerically; a solution found starting from \eqref{SUSYBBuv} with arbitrary values of $c_+$ and $c_-$ will generically be singular in the IR, with a divergent Kretschmann scalar \cite{Elander:2011mh}.

\subsection{Exploring the baryonic branch}
\label{h1Dep}
We saw in section \ref{SUSYsols} that, constrained by the requirement of regularity, and ignoring the possible shift of the dilaton, the SUSY solutions form a one dimensional family. It is convenient to parametrise the solutions either by $h_1$, which is defined by the IR expansions \eqref{SUSYBBir}, or by $c_+$, which is defined by the UV expansions \eqref{SUSYBBuv}. The relationship between $h_1$ and $c_+$ is known only numerically, but for these SUSY solutions we have
\begin{align}
	c_+ \sim \frac{ 3^{1/3} h_1}{4}			\label{cph1}
\end{align}
for large values of $c_+$ and $h_1$ \cite{Gaillard:2010qg}. As we will see in section \ref{fieldTheory}, in the rotated solutions $h_1$ and $c_+$ correspond to the parameter which explores the baryonic branch; we recover the KS solution \cite{Klebanov:2000hb} itself in the limit $h_1,c_+ \to \infty$. We postpone further discussion of this limit until section \ref{DKlimit}, where we consider its non-SUSY generalisation \cite{Dymarsky:2011ve}.

\label{MNlimit} Taking the opposite limit, $c_+\to0$, we find that $h_1\to 2\Nc$. This corresponds to the Chamseddine-Volkov/Maldacena-N\'u\~nez (CVMN) solution \cite{Maldacena:2000yy,Chamseddine:1997nm}. This is considerably simpler than the general case, and exact expressions are known for the functions which describe the solution:
\begin{align}
	\frac{e^{2g}}{\Nc} = \frac{e^{2k}}{\Nc} &= 1																,		&
	\frac{e^{2h}}{\Nc} &=  \r \coth 2\r - \frac{\r^2}{\sinh^2 2\r} - \frac14			,		\nn\\
																		a = b &= \frac{2\r}{\sinh{2\r}	}					,		&
												e^{4\Phi-4\phi_0} &= \frac{\Nc}{4}e^{-2h} \sinh^2 2\r	.				\label{MNexact}
\end{align}
Note that while the IR can be obtained simply by setting $h_1=2\Nc$ in \eqref{SUSYBBir},
\begin{align}
	\frac{e^{2h}}{\Nc} = \r^2 - \frac49 \r^4 + O(\r^6),	\quad
	a									 = 1 -\frac23 \r^2 + O(\r^4)		,		\quad
	e^{4\Phi-4\phi_0}	 = 1+ \frac{16}{9} \r^2 + O(\r^4) ,					\label{MNir}
\end{align}
the UV is qualitatively different from the general case:
\begin{align}
	\frac{e^{2h}}{\Nc} = \r - \frac14 + O\bigl(e^{-4\r}\bigr),	\qquad
	a									 = 4\r e^{-2\r} + O\bigl(e^{-6\r}\bigr),	\qquad
	\Phi	  	 = \r + O(\log\r)						.										 					\label{MNuv}
\end{align}
Of particular significance here is the fact that the dilaton grows without bound in the UV. As anticipated above, this means that we cannot apply the rotation procedure \eqref{rotationhh}.

\begin{figure}[htbp]
	\centering
		\includegraphics{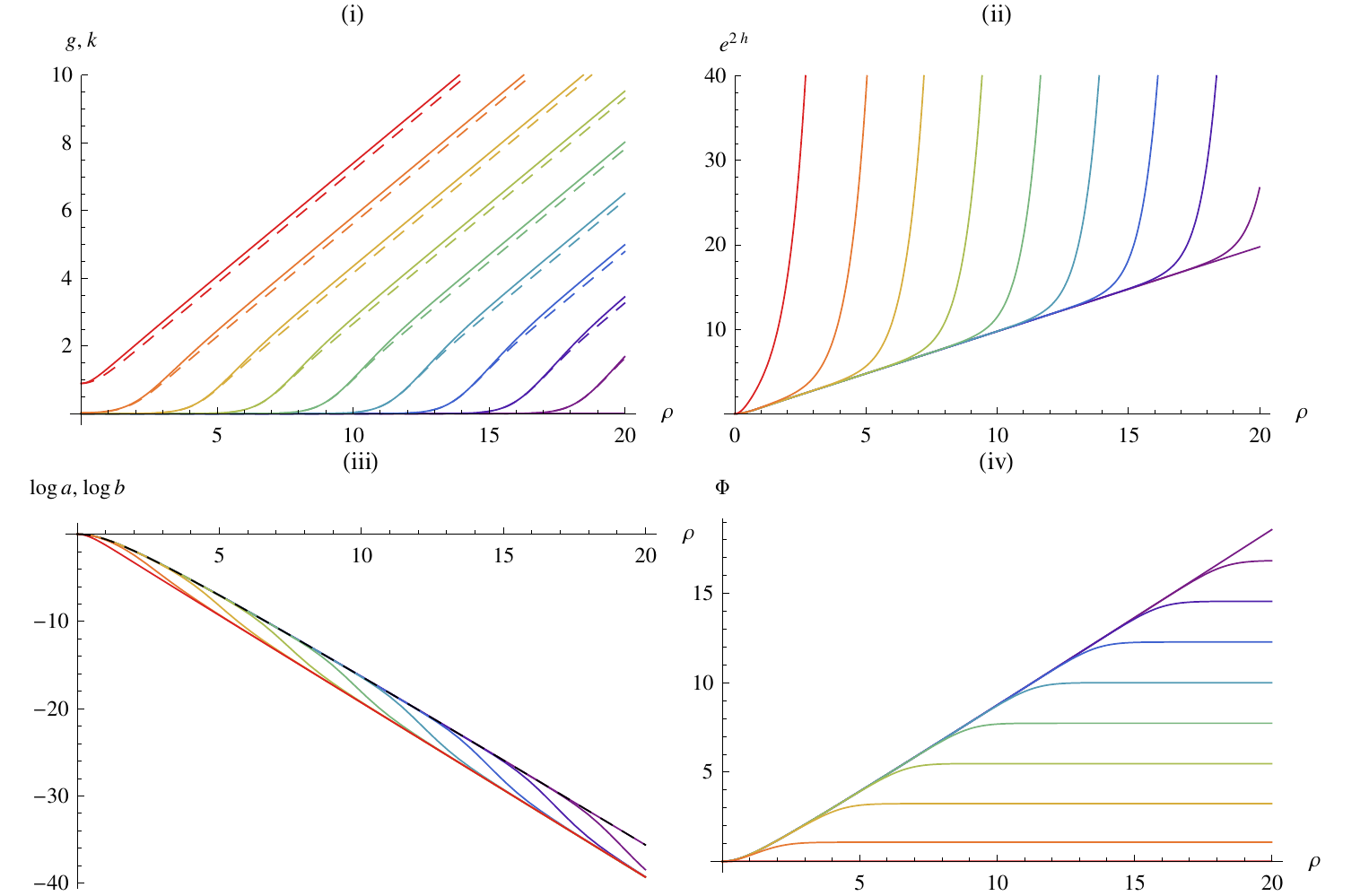}
	\caption{ Plots of (i) $g$ (solid) and $k$ (dashed), (ii) $e^{2h}$, (iii) $\log a$ (solid) and $\log b$ (dashed, black), and (iv) $\Phi$, for the SUSY solutions with $2\le h_1 \le 12$, increasing from purple to red. Here we set $\Nc=1$ and $\phi_0 = 0$.}
	\label{plotSUSY}
\end{figure}

\begin{figure}[htbp]
	\centering
		\includegraphics{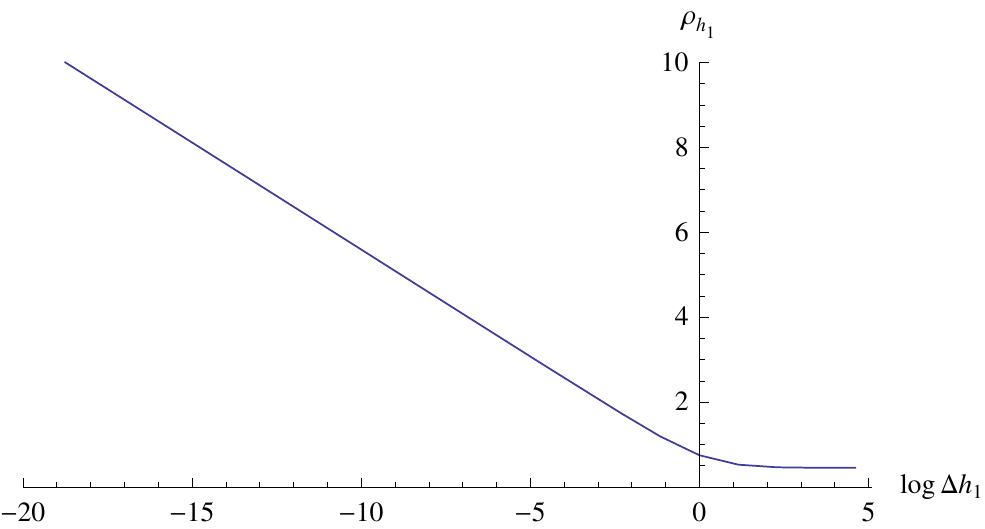}
	\caption{ Plot showing the dependence of $\rMN$ on $\Delta h_1 = h_1-2\Nc$ in the SUSY solutions. For the purposes of this plot we define $\rMN$ by $k'(\rMN)=1/3$, corresponding to the transition between the CVMN UV ($k=\text{constant}$) and the generic UV ($k\sim 2\r/3$).}
	\label{plotRhoMN}
\end{figure}

In the general case $h_1>2\Nc$ the system follows the CVMN solution closely, before switching to the generic UV \eqref{SUSYBBuv} for large $\r$ (figure \ref{plotSUSY}). That is, we can identify a scale $\rMN$ below which \eqref{MNexact} is almost satisfied, and above which $g$, $h$, and $k$ grow exponentially, $\Phi$ quickly goes to a constant, and $a\ne b$. Notice that $b$ is completely unaffected by this; the exact result $b=2\r/\sinh2\r$ holds for all $h_1$. As $h_1$ is increased, $\rMN$ moves further into the IR (figure \ref{plotRhoMN}).

We noted above that the rotation procedure could not be applied to the CVMN solution \eqrefs{MNexact}{MNuv}. Specifically, in section \ref{SUSYoverview} we chose a particular value for the constant appearing in the warp factor $\hat{h}=1-\kappa^2e^{2\Phi}$, namely $\kappa = e^{-\Phi_\infty}$. In the CVMN solution $\Phi$ grows without bound and this identification is no longer possible. Nevertheless, it turns out that there is a sense in which we do obtain the (unrotated) CVMN solution by taking the limit $h_1\to2\Nc$ in the (rotated) baryonic branch. To see this, note that as we take the limit $h_1\to2N_c$, we find that $\Phi_\infty\to\infty$, and so $\kappa\to0$. In this limit we see that $\hat{h}\to1$ and the additional fields in \eqref{fieldsRot} vanish, returning us to the unrotated system \eqrefs{vielbeinUnrot}{fieldsUnrot} at any finite $\r$.

More explicitly, in a generic solution on the baryonic branch, the dilaton becomes almost constant approximately at the scale $\rMN$ (figure \ref{plotSUSY}). Provided $\rMN$ is large enough that the UV expansions are valid, we see from \eqref{MNuv} that for $\r<\rMN$ we have $\Phi\sim\r$. Taken together, these observations mean that we can write $\Phi_\infty \sim \rMN$. We then find numerically (figure \ref{plotRhoMN}) that $\k^2\sim e^{-2\rMN}\sim h_1-2\Nc\to0$ for $h_1\to2\Nc$. 

In effect, taking the limit $h_1\to2\Nc$ in the rotated solutions simply pushes the scale $\rMN$ to infinity, while in the region $\r<\rMN$ the solution becomes exactly the CVMN one. However, it is important to note that the two cases are qualitatively different, and the limit is not entirely smooth. In particular, we can expect any quantity which depends on the the UV asymptotics of the background to behave discontinuously as we take the limit.

\section{Breaking SUSY}\label{breakingSUSYsection}

\subsection{Deformation of $h_1=2N_2$ case}\label{GTVlimit}
The CVMN solution \cite{Maldacena:2000yy,Chamseddine:1997nm} which we obtain in the limit $h_1=2N_c$ (section \ref{MNlimit}) can be described in terms of $SO(4)$ gauged seven-dimensional supergravity. The $SO(4)$ gauge group corresponds in the full ten-dimensional description to rotations of the 3-sphere $(\tht,\pht,\psi)$.
	
In order to get a four-dimensional world-volume theory we wrap 5-branes on the $S^2$ $(\theta,\vph)$. There is no covariantly constant spinor on $S^2$, so to preserve some  supersymmetry we have to turn on a gauge field so as to cancel the spin connection of the $S^2$ in the variation of a fermion:
\begin{align}
	\delta\Psi \sim D_\mu\epsilon = (\partial_\mu + \omega_\mu^{\nu\r}\gamma^{\nu\rho} - A_\mu^{ij}\Gamma^{ij} )\epsilon.
\end{align}
This can be achieved, preserving $\N=1$ SUSY, with an abelian field $U(1)\subset SU(2)_L$, where $SO(4) \sim SU(2)_R \times SU(2)_L $.\footnote{Alternatively, we could preserve $\N=2$ SUSY by choosing the $U(1)$ to be in a diagonal $SU(2)_D \subset SU(2)_R \times SU(2)_L$, as in \cite{Bigazzi:2001aj,Gauntlett:2001ur,Andrews:2006aw}.	} In the ten-dimensional description, this corresponds to the `twist' given by the mixing with the $S^2$ coordinates $\th$ and $\vph$ in \eqref{vielbeinUnrot},	
\begin{align}
	E^3 \sim \tilde\omega_3 + \cos\th d\vph.
\end{align}
The resulting solution is singular in the IR. However, we can obtain the regular CVMN solution by allowing a non-abelian $SU(2)$ field. In the ten-dimensional description this shows up in the additional mixing parametrised by $a(\r)$ in \eqref{vielbeinUnrot}. When $a(\r)=1$, as occurs for instance at the origin in the SUSY solution, the gauge field is pure gauge; the gauge transformation which removes the field can be written as a coordinate transformation which removes the mixing \cite{Gimon:2002nr,Gursoy:2005cn}.

This solution was generalised in \cite{Apreda:2003gc} by solving the full equations of motion rather than the BPS equations, and by allowing a full $SO(4)$ gauge field. We are interested here in the simplest SUSY-breaking deformation of the CVMN solution, where we keep the $SU(2)$ gauge group, and introduce a mass term which breaks SUSY. This corresponds to the globally regular extremal solutions obtained by Gubser, Tseytlin and Volkov (GTV) \cite{Gubser2001}.

For these non-SUSY solutions we no longer have an exact solution as in \eqref{MNexact}, although we still have
\begin{align}
	\frac{e^{2g}}{\Nc} = \frac{e^{2k}}{\Nc} = 1,							\qquad\qquad
	a = b						\label{GTVsimple}
\end{align}
for all $\r$. Instead we must rely on expansions in the IR and UV. In the IR, we have qualitatively the same as in \eqref{MNir}:
\begin{align}
	\frac{e^{2h}}{\Nc} = \r^2 - &\left( \frac29 + \frac{v_2^2}{2} \right) \r^4 + O(\r^6),	\qquad\qquad
	a									 = 1 		+	v_2  \r^2 + O(\r^4)		,		\nn\\
&	e^{4\Phi-4\phi_0}	 = 1+ \left( \frac43 + v_2^2 \right) \r^2 + O(\r^4), 					\label{GTVir}
\end{align}
where we have introduced $v_2$ to parametrise the SUSY-breaking deformation. Comparing to \eqref{MNir} we see that setting $v_2 = -2/3$ recovers the SUSY CVMN solution. 

As explained in \cite{Gubser2001}, to obtain a regular UV we need $-2 \le v_2 \le 0$. We then obtain substantially different behaviour to that in the SUSY case. Adapting the notation of \cite{Apreda:2003gc},
\begin{align}
	\frac{e^{2h}}{\Nc} = \r + G_\infty + O\left(\frac{1}{\r}\right) ,	\qquad
	a									 = M_a \r^{-1/2} + O( \r^{-3/2} ),	\qquad
	\Phi	  	 = \r + O(\log\r),																 					\label{GTVuv}
\end{align}
where the parameters $M_a$ and $G_\infty$ can be considered functions of $v_2$. The main qualitative difference is the presence of additional terms decaying slower than exponentially in the expansions for $e^{2h}$ and $a$. This is interpreted in \cite{Apreda:2003gc} as corresponding to a mass which breaks SUSY.

\begin{figure}[htbp]
	\centering
		\includegraphics{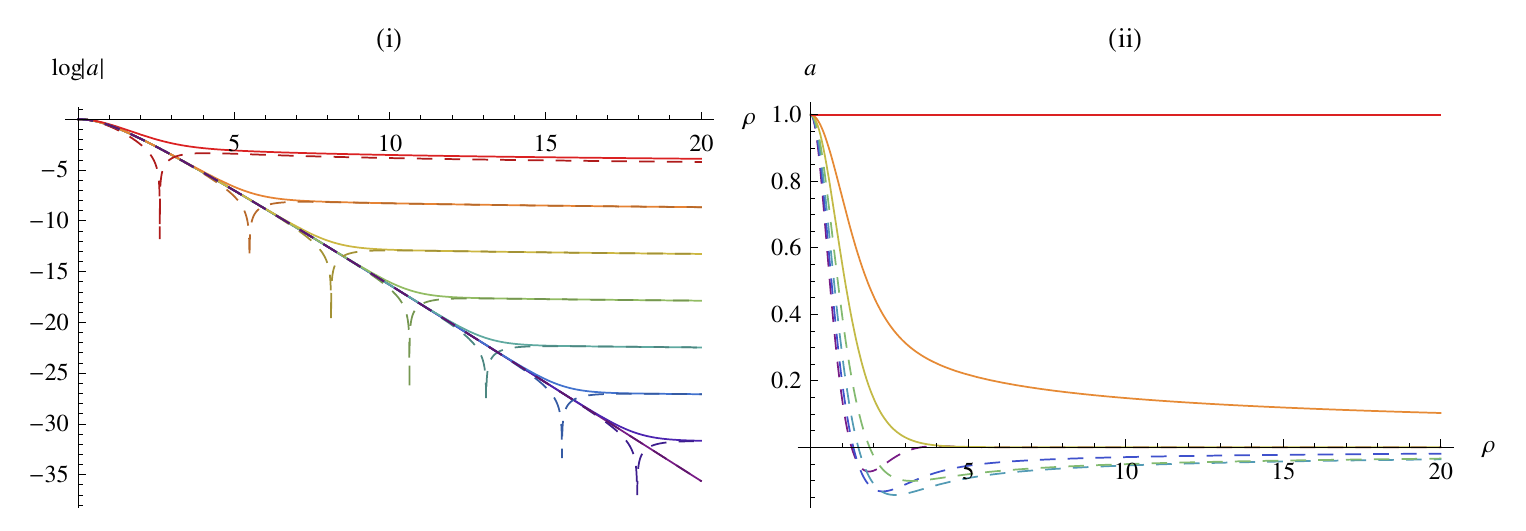}
	\caption{(i) Plot of $\log a$ against $\r$ for $-1/10 \le \Delta v_2 \le 1/10$, where $\Delta v_2 = v_2 + 2/3$, showing the transition between 
$a\sim e^{-2\r}$ and $a\sim\sqrt{\r}$ at $\r\sim\rSUSY$. The dashed curves correspond to $v_2<-2/3$, for which $a=0$ at $\rSUSY$.	\newline
(ii)  Plot of $a$ against $\r$ for the full range $-2\le v_2\le 0$. Again, the dashed curves correspond to $v_2<-2/3$, for which $a$ has at least one zero.	The additional oscillations which are in fact present in the case $v_2=-2$ (purple) are not visible at this scale.}
	\label{plotGTVa}
\end{figure}

\begin{figure}[htbp]
	\centering
		\includegraphics{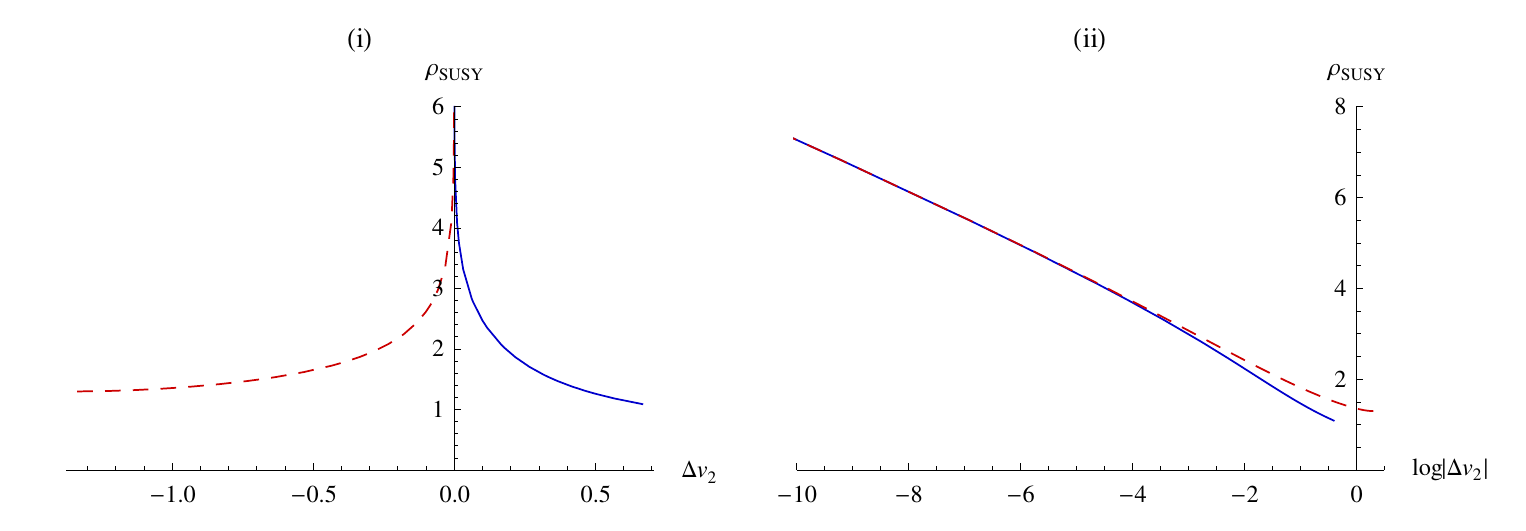}
	\caption{ Plots showing the dependence of $\rSUSY$ on $\Delta v_2 = v_2+2/3$. For the purposes of this plot we define $\rSUSY$ by $\left|\Delta a\right| = a_\text{SUSY}$. The solid blue curve corresponds to $v_2>-2/3$ and the dashed red curve to $v_2<-2/3$.}
	\label{plotRhoSUSY}
\end{figure}

The effect of the SUSY-breaking deformation is most clearly understood by considering $a$, which is affected at leading order (figure~\ref{plotGTVa}~(i)). We see that the non-SUSY solutions are characterised by a scale $\rSUSY$. For $\r<\rSUSY$, the qualitative behaviour is that of the SUSY solution, \eqrefs{MNexact}{MNuv}, while for $\r>\rSUSY$ the non-SUSY UV of \eqref{GTVuv} takes over. For a generic non-SUSY solution we can define the deformation to $a$ as
\begin{align}
	\Delta a = a - a_\text{SUSY}.
\end{align}
Then we can think of $\rSUSY$ as the scale at which the deformation $\Delta a$, which decays slowly in the UV, is of comparable magnitude to $a_\text{SUSY}$, which decays much faster. As a result, $\rSUSY$ moves towards the IR as we move further from the SUSY solution (figure \ref{plotRhoSUSY}). Note that this does not relate in a obvious way to the SUSY-breaking scale, which it would be more natural to associate with the scale above which $\Delta a$ has decayed significantly, and which moves into the UV as we move further from the SUSY solution.

For $v_2\ge0$, $a$ is always positive, and for $v_2=0$, $a=1$ for all $\r$. As noted above, this means that the gauge field is pure gauge, and we can remove the mixing between the spheres by a change of coordinates. Thus in this case the internal geometry is simply $S^2\times S^3$. Aside from the behavior of $a$, the UV is otherwise unchanged --- the other functions $h$ and $\Phi$ still behave according to \eqref{GTVuv}.

\begin{figure}[htbp]
	\centering
		\includegraphics{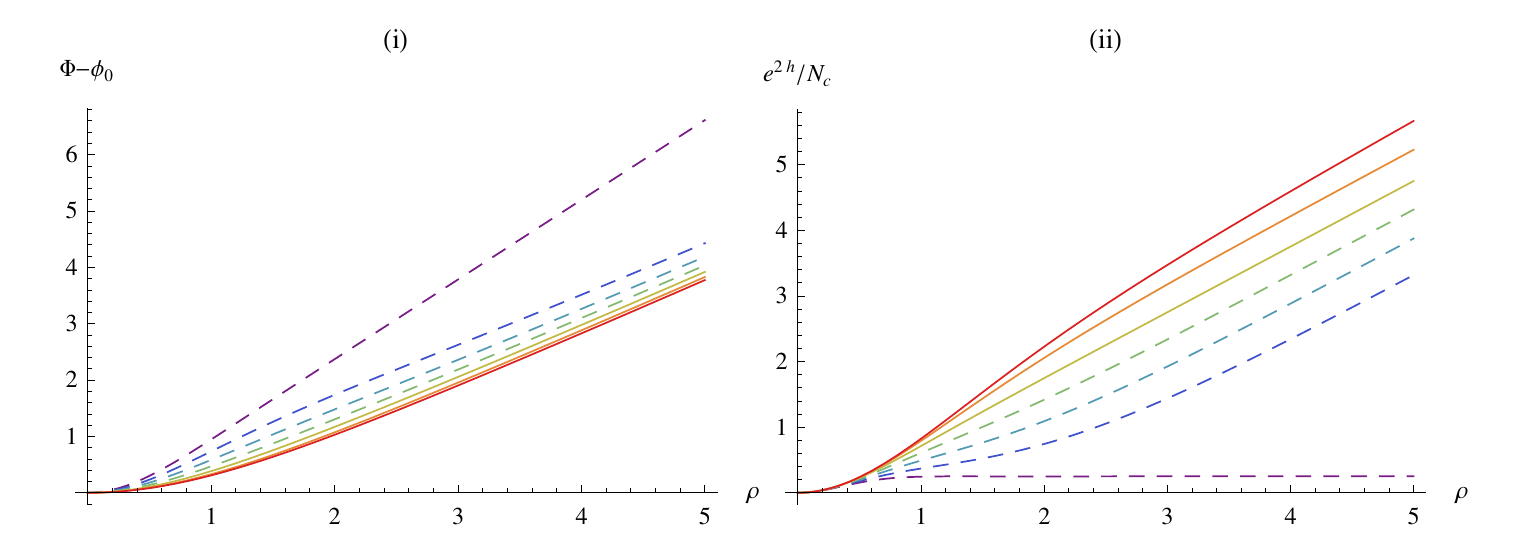}
	\caption{Plots of (i) $\Phi$ and (ii) $e^{2h}$ against $\r$, for $-2\le v_2<-2/3$ (dashed curves) and $-2/3\le v_2\le0$ (solid curves), showing the difference between the generic UV \eqref{GTVuv} and the limiting case \eqref{GTVminuv}. }
	\label{plotGTVhphi}
\end{figure}

For $v_2<-2/3$, $a$ has at least one zero. As $v_2$ is reduced, $a$ picks up more oscillations, and in the limiting case there are infinitely many zeros. In this limit the UV of the other functions no longer that of \eqref{GTVuv} (see figure \ref{plotGTVhphi}). Instead the system approaches the `special Abelian solution' of \cite{Gubser2001}; 
\begin{align}
	\frac{e^{2h}}{\Nc} \to \frac{1}{4},   \qquad\qquad    \Phi \to \sqrt{2}\r.		\label{GTVminuv}
\end{align}

\subsection{Deformation of general case}\label{generalNonSUSY}
We now turn to the solution which was originally presented in \cite{Bennett:2011va}. The aim there was to find a non-supersymmetric generalisation, preserving the symmetries and structure of the baryonic branch solutions with $h_1\ne2\Nc$ discussed in section \ref{SUSYsection}. This is analogous to the way in which the GTV solutions (section \ref{GTVlimit}) generalise the CVMN solution \eqref{MNexact}.

In the non-SUSY case we can no longer make use of the master equation \eqref{masterEq}, which was derived from the BPS equations. Instead we have to solve the full Einstein, Maxwell, dilaton and Bianchi equations of the system. This amounts to a system of six coupled non-linear second-order equations, together with a first-order Hamiltonian constraint. These are included in appendix \ref{EOMappendix}. We look for solutions to these equations in the form of IR and UV expansions with similar forms to the SUSY case \eqrefs{SUSYBBir}{SUSYBBuv}.

In the IR we simply impose that the solution is regular, and that the 2-sphere shrinks to zero radius at $\r=0$, as in \eqref{SUSYBBir}. We then have expansions of the form
\begin{align}
e^{2g}		&= \sum_{n=0}^{\infty}g_n \r^n,			&
e^{2h}		&= \sum_{n=2}^{\infty}h_n \r^n,			&
e^{2k}		&= \sum_{n=0}^{\infty}k_n \r^n,			\nn	\\
e^{4\Phi}	&= \sum_{n=0}^{\infty}f_n \r^n,			&
a					&= \sum_{n=0}^{\infty}w_n \r^n,			&
b					&= \sum_{n=0}^{\infty}v_n \r^n.							\label{nonsusyAnsatzIR}
\end{align}
Substituting into the equations of motion \eqrefs{eq:eomg}{eq:eomb} we find five independent parameters, which we take to be $k_0$, $f_0$, $k_2$, $v_2$ and $w_2$. We relabel $k_0= h_1/2$ and $f_0 =e^{4\phi_0}$, so that we can recover the SUSY solution \eqref{SUSYBBir} by setting
\begin{align}
	k_2 = \frac{2h_1}{5} - \frac{8\Nc^2}{h_1}	,	\qquad
	v_2 = - \frac23														,	\qquad
	w_2 = \frac{8\Nc}{3h_1} - 2.											\label{susyValsIR}
\end{align}
After the relabeling, the five independent parameters are\footnote{Notice that $h_1$ does \emph{not} refer to the coefficient of $\r$ in the expansion for $e^{2h}$, as would be expected from the form of \eqref{nonsusyAnsatzIR}. This unfortunate notation should not cause confusion because that term will always be zero to ensure regularity.} 
\begin{align}
	h_1,	\qquad	\phi_0,	\qquad	k_2,	\qquad	v_2,	\qquad w_2,			\label{IRparam}
\end{align}
and the expansions are qualitatively the same as the SUSY case \eqref{SUSYBBir}:\footnote{More complete expressions, both for the IR expansions here and the UV \eqref{nonsusyExpUV}, can be found in an appendix of \cite{Bennett:2011va}.}
\begin{align}
e^{2g}		&= \frac{h_1}{2}	+		\frac{h_1}{2}\left(	1 - \frac{k_2}{h_1}
																											-	\frac{4\Nc^2}{h_1^2}
																											-	\frac{\Nc^2 v_2^2}{h_1^2}
																											+	\frac{w_2^2}{4}						\right)\r^2	+	O(\r^4),			\nn\\
e^{2h}		&= \frac{h_1}{2}\r^2-	\frac{h_1}{6}\left(	1	-	\frac{2k_2}{h_1}
																											-	\frac{4\Nc^2}{3h_1^2}
																											+ \frac{3\Nc^2 v_2^2}{h_1^2}
																											+	\frac{3w_2^2}{4}					\right)\r^4 + O(\r^6),			\nn\\
e^{2k}		&= \frac{h_1}{2}	+		k_2	\r^2		+	O(\r^4)																									 ,			\qquad\qquad
e^{\Phi-\phi_0}
					= 1							+		\frac{\Nc^2}{h_1^2}	\left(	\frac43 + v_2^2			\right)\r^2		+ O(\r^4),			\nn\\
a					&= 1	+	w_2	\r^2	+	O(\r^4),			\qquad\qquad
b					=	1		+	v_2	\r^2	+	O(\r^4).							\label{nonsusyExpIR}
\end{align}

In the UV we use a particular generalisation of the SUSY solutions \eqref{SUSYBBuv}:
\begin{align}
e^{2g}		&= \sum_{m=0}^\infty \sum_{n=0}^{m}	G_{mn} 		\r^n e^{4(1-m)\r/3},		&
e^{2h}		&= \sum_{m=0}^\infty \sum_{n=0}^{m}	H_{mn} 		\r^n e^{4(1-m)\r/3},		\nn\\
e^{2k}		&= \sum_{m=0}^\infty \sum_{n=0}^{m} K_{mn} 		\r^n e^{4(1-m)\r/3},	&
e^{4\Phi} &= \sum_{m=1}^\infty \sum_{n=0}^{m}	\Phi_{mn} \r^n e^{4(1-m)\r/3},		\nn\\
a					&= \sum_{m=1}^\infty \sum_{n=0}^{m}	W_{mn} 		\r^n e^{2(1-m)\r/3},		&
b					&= \sum_{m=1}^\infty \sum_{n=0}^{m}	V_{mn} 		\r^n e^{2(1-m)\r/3}.					\label{nonsusyAnsatzUV}
\end{align}
This particular ansatz will not be sufficient to include all cases; for example we have seen in section \ref{GTVlimit} that the GTV solutions with $h_1=2\Nc$ have a completely different form \eqref{GTVuv} in the UV. We will find other limits in which this is the case, and which will have to be treated separately.

As in the IR, we substitute the ansatz \eqref{nonsusyAnsatzUV} into the equations of motion, and in this case we find nine independent parameters, which we take to be
\begin{align}
K_{00},\quad K_{30},\quad H_{10},\quad H_{11},\quad\Phi_{10},\quad \Phi_{30},\quad W_{20},\quad W_{40},\quad V_{40},
\end{align}
and which we again relabel to make contact with the SUSY case:
\begin{align}
	K_{00} 		= \frac{2c_+}{3},	\quad
	H_{10}		=	\frac{Q_o}{4},	\quad
	\Phi_{10} = e^{4\Phi_\infty}, \quad
	K_{30}		=	\frac{c_- -64e^{4\rho_o}c_+^3}{48c_+^2},	\quad
	W_{40} 		= 2e^{\rho_o}.
\end{align}

The nine relabeled independent parameters are then
\begin{align}
c_+ ,\quad c_-,\quad \Phi_\infty ,\quad Q_o ,\quad \rho_o ,\quad  H_{11} ,\quad W_{20} ,\quad \Phi_{30} ,\quad  V_{40},
\label{UVparam}
\end{align} 
and the expansions are
\begin{align}
e^{2g}		&= c_+	e^{\frac43\r}						-	( 4H_{11} \r + Q_o + 2c_+ W_{20}^2	)			+		O\bigl( e^{-\frac43\r} \bigr)			,			\quad
e^{2h}		 = \frac{c_+}{4} e^{\frac43\r}	+				H_{11}\r  + \frac{Q_o}{4}  					+		O\bigl( e^{-\frac43\r} \bigr)			,			\nn\\
e^{2k}		&= \frac{2c_+}{3}e^{\frac43\r}	+					\frac{c_+ W_{20}^2}{3}						+		O\bigl( e^{-\frac43\r} \bigr)  		,			\quad
e^{\Phi-\Phi_\infty}	 
					= 1							-		\left(	\frac{3\Nc^2 }{2c_+^2}\r		-	e^{-4\Phi_\infty} \frac{\Phi_{30}}{4}	\right) e^{-\frac83\r}
																																									+		O\bigl( e^{-4\r  } \bigr)  		,			\nn\\
a					&= W_{20} e^{-\frac23\r}	+	\left[			\left(	\frac{3H_{11}W_{20}}{c_+}
																																	+ \frac{10W_{20}^3}{3}		\right)\r	
																									+ 2e^{2\r_o}																					\right]e^{-2\r}
																																									+		O\bigl( e^{-\frac{10}{3}\r} \bigr)  	,			\label{nonsusyExpUV}\\
b					&= \frac{9W_{20}}{4} e^{-\frac23\r}
																+ \left[		 	\frac{10W_{20}^3}{3}	\r^2		
																						+ \left(	4e^{2\r_o}
																																	-	\frac{Q_o W_{20}}{c_+}
																																	-	\frac{23W_{20}^3}{6}	\right)\r
																						+	V_{40}										\right]e^{-2\r}
																																									+		O\bigl( e^{-\frac{10}{3}\r} \bigr)  	.			\nn
\end{align}
The most significant difference here when compared to the SUSY expansions \eqref{SUSYBBuv} is the presence of the new terms at leading order in the UV in $a$ and $b$. This corresponds to the presence of the additional terms proportional to $\r^{-1/2}$ in $a$ which we saw in the GTV solutions \eqref{GTVuv}, and we will see in section \ref{fieldTheory} that the interpretation as a mass term still applies. The fact that the extra terms we obtain here are exponential rather than polynomial in $\r$ is related to the qualitatively different UV asymptotics in the baryonic branch \eqref{SUSYBBuv} as opposed to the CVMN solution \eqref{MNuv}.

We can recover the SUSY case from \eqref{nonsusyExpUV} by setting
\begin{align}
	H_{11}=\frac{\Nc}{2}	,\quad
	W_{20}=0				,\quad
	\Phi_{30}= - \frac{3\Nc}{4c_+^2}e^{4\Phi_\infty} (3\Nc+4Q_o)		,\quad
	V_{40}=\frac{2}{\Nc} e^{2\rho_o} (\Nc+Q_o).
\label{recoverSUSYUV}
\end{align}
For the regular SUSY solution \eqrefc{SUSYBBir}{SUSYBBuv} we also need $\r_o = 0$ and $Q_o = -\Nc$.

In summary, our solutions are described by the fourteen parameters: the five from the IR \eqref{IRparam} and nine from the UV \eqref{UVparam}. However, if we consider only solutions which match both the IR and UV expansions \eqrefc{nonsusyAnsatzIR}{nonsusyAnsatzUV} these are clearly not all independent. There can be at most five independent parameters, as the required solutions can be parametrised by the IR boundary conditions alone. However, we generically expect even fewer.

Our goal is to find a solution which smoothly interpolates between the IR and UV expansions. This will require that these two parametrisations lead to identical functions. We can express this as a system of twelve equations \footnote{We write the functions resulting from a given choice of the IR parameters $\{h_1, k_2, v_2, w_2\}$ in the form $g(h_1,k_2,v_2,w_2;\r)$. Similarly the expressions of the form $g(c_+,c_-,Q_o,\r_o,H_{11},W_{20},\Phi_{30},V_{40};\r)$ refer to the functions resulting from a given choice of the UV parameters.},	
\begin{align}
\begin{tabular}{r @{\ } c @{\ } l		@{\qquad} 	 	r @{\ } c @{\ } l}
	$									g(h_1 \dots w_2;\rho)		$	&	$=			$	&	$								g(c_+ \dots  V_{40};\rho)	$,	&				
	$		\frac{d}{d\r}	g(h_1 \dots w_2;\rho)		$	&	$=			$	&	$\frac{d}{d\r}	g(c_+ \dots  V_{40};\rho)	$,	\\
	$									h(h_1 \dots w_2;\rho)		$	&	$=			$	&	$								h(c_+ \dots  V_{40};\rho)	$,	&				
	$		\frac{d}{d\r}	h(h_1 \dots w_2;\rho)		$	&	$=			$	&	$\frac{d}{d\r}	h(c_+ \dots  V_{40};\rho)	$,	\\
																							&	$\vdots	$	&																							 		&				
																							&	$\vdots	$	&																							 		\\
	$									b(h_1 \dots w_2;\rho)		$	&	$=			$	&	$								b(c_+ \dots  V_{40};\rho)	$,	&				
	$		\frac{d}{d\r}	b(h_1 \dots w_2;\rho)		$	&	$=			$	&	$\frac{d}{d\r}	b(c_+ \dots  V_{40};\rho)	$.
\end{tabular}	\label{matchingEq}
\end{align}
This system can be further reduced using the constraint (appendix \ref{EOMappendix}). This means we can for instance express the derivative of one of the functions in terms of the other functions and their derivatives. This leaves us with a system of eleven independent equations which we would expect to allow us to solve for eleven of our fourteen parameters. Although in principle further redundancy in the system of equations \eqref{matchingEq} would allow for more independent parameters up to a maximum of five, the numerical analysis discussed in \cite{Bennett:2011va} and below appears to support this conclusion. Of the three remaining parameters, one corresponds to our ability to shift the dilaton, which has no other effect on the solution. The final two parameters we then associate with movement along the baryonic branch and finally the breaking of SUSY.

In much of the following it will be convenient to describe the solution space in terms of the parameters that appear in the IR expansions. Firstly, as in \cite{Bennett:2011va} the smaller number of parameters makes finding suitable numerical solutions much simpler starting from the IR. Secondly, our IR ansatz \eqref{nonsusyAnsatzIR} imposes a comparatively natural restriction on the solutions, while the UV ansatz \eqref{nonsusyAnsatzUV} is more arbitrary, merely being a plausible candidate for a generalisation of the most usual SUSY solution. Indeed, as discussed we know that it does not apply in several interesting special cases.

\begin{figure}[htbp]
	\centering
		\includegraphics{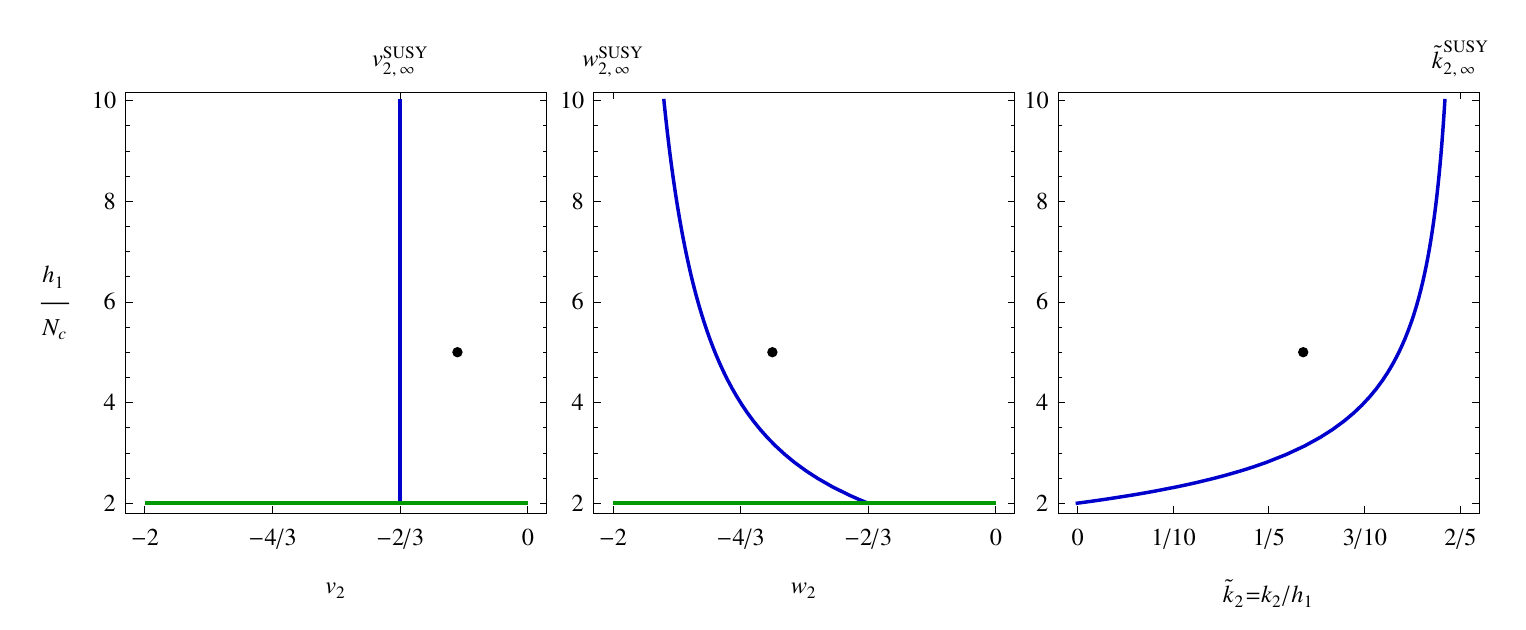}
	\caption{The space of solutions, seen in terms of $(v_2,h_1)$, $(w_2,h_1)$, and $(\tilde k_2,h_1)$, where $\tilde k_2 = k_2/h_1$. The blue curves denote the SUSY baryonic branch (section \ref{SUSYsection}), and the green lines correspond to the GTV solutions (section \ref{GTVlimit}). Note that all the GTV solutions have $k_2=0$, as can be seen from \eqref{GTVsimple}. Any solution to the equations of motion of the form \eqref{nonsusyAnsatzIR} is represented by a point on each of these diagrams. If we require a well-behaved UV, specifying the position on one diagram is sufficient to determine the positions on the other two. For example the marked point represents schematically a generic solution of the sort presented in \cite{Bennett:2011va}. The values marked at the top show the SUSY values in the limit $h_1\to\infty$, corresponding to the KS solution \cite{Klebanov:2000hb}.
	}
	\label{plotParamSpaceSimple}
\end{figure}

To allow contact with the SUSY case, we choose $h_1$ to parametrise the position along the baryonic branch. We could then in principle choose any combination of the remaining IR parameters $v_2$, $w_2$ and $k_2$ to describe the remaining degree of freedom (figure \ref{plotParamSpaceSimple}). It turns out that a description in terms of $v_2$ is usually simplest; we see from \eqref{susyValsIR} that its SUSY value, $v_2^\text{SUSY}=-2/3$, is independent of $h_1$. 

\subsection{Finding globally regular solutions}\label{tuningSection}
In order for us to be able to conclude that the IR expansions of the form \eqref{nonsusyAnsatzIR} and the UV expansions of the form \eqref{nonsusyAnsatzUV} describe the same system of solutions, it is necessary to find numerical solutions interpolating between them. This was achieved in \cite{Bennett:2011va} for isolated examples, simply by manually searching the IR parameter space for solutions with the expected UV behaviour. However, without having a good understanding of the structure of the parameter space it was difficult to make progress. In particular, the approach was in practice limited to solutions very close to the SUSY case (i.e. $v_2\approx -2/3$).

Fortunately, we can make use of the simpler system of GTV solutions (section \ref{GTVlimit}). Just as the CVMN solution \eqref{MNexact} can be obtained from the SUSY baryonic branch solution \eqrefs{SUSYBBuv}{SUSYBBir} in the limit $h_1\to2\Nc$, we would expect to obtain the GTV solutions from our non-SUSY generalisation of the baryonic branch in the same limit. In the IR, this is indeed the case; by setting $w_2=v_2$ and $k_2=0$ in our solution we recover \eqref{GTVir}. Of course, there is no way to obtain the GTV UV \eqref{GTVuv} from UV expansions of the form \eqref{nonsusyAnsatzUV}, but this is to be expected given that the equivalent statement is also true in the SUSY case --- the CVMN UV \eqref{MNuv} cannot be obtained as a simple limit of the generic UV \eqref{SUSYBBuv}.

As the GTV system has no redundant parameters in the IR, it is simple to generate numerical solutions. It is then possible to deform this well-understood case by increasing $h_1$ and adjusting $w_2$ and $v_2$ slightly to correct the UV behaviour. More precisely, for a given value of $v_2=\Delta v_2 -2/3$ it is trivial to obtain a numerical solution with $h_1=2\Nc$, for which $w_2=v_2$ and $k_2=0$. We then deform this by keeping $\Delta v_2$ fixed and setting $h_1 = 2\Nc+\Delta h_1$. If we use a small perturbation $\Delta h_1$, we will require corrections of the form
\begin{align}
	w_2 = w_2^\text{SUSY}(\Delta h_1) + \Delta v_2 + \delta w_2(\Delta h_1,\Delta v_2),		\qquad 
	k_2 = k_2^\text{SUSY}(\Delta h_1) + \delta k_2(\Delta h_1,\Delta v_2),
\end{align}
where $\delta w_2$ and $\delta k_2$ are extremely small.

The far UV of the solutions obtained in this way match our general ansatz \eqref{nonsusyAnsatzUV}, justifying our assumption that the GTV solutions can be viewed as a limit of our deformations of the general case.

In itself, this yields a considerable advance over using only the approach described in \cite{Bennett:2011va} --- it gives us access to solutions with $h_1\approx 2\Nc$ and general $v_2$, in addition to those with $v_2\approx-2/3$ and general $h_1$. More significantly, however, it allows us to understand the behaviour of solutions with generic values of both $h_1$ and $v_2$ in terms of the corresponding solutions in the two limits.


\section{The two-dimensional solution space}\label{2dSS}
\subsection{Combining the effects of $h_1$ and $v_2$}\label{h1v2dep}
As we have seen in section \ref{generalNonSUSY}, the system is described by a two-dimensional parameter space, corresponding to the position along the baryonic branch and the size of the SUSY-breaking deformation. We generate numerical solutions starting from the IR, so we are led to the choice of $h_1$ and one of the three SUSY-breaking parameters $\{w_2, v_2, k_2\}$. Of these $v_2$ turns out to be most convenient because $v_2^\text{SUSY}$ is independent of $h_1$. 

In section \ref{h1Dep} we described the effect of varying $h_1$ in terms of the scale $\rMN$, corresponding to the transition between the CVMN behaviour \eqrefs{MNexact}{MNuv} and the generic (KS-like) behaviour \eqref{SUSYBBuv}. Similarly, in section \ref{GTVlimit} we introduced the scale $\rSUSY$, associated with the transition between the qualitatively SUSY CVMN behaviour and the (non-SUSY) GTV UV \eqref{GTVuv}.

\begin{figure}[htb]
	\centering
		\includegraphics{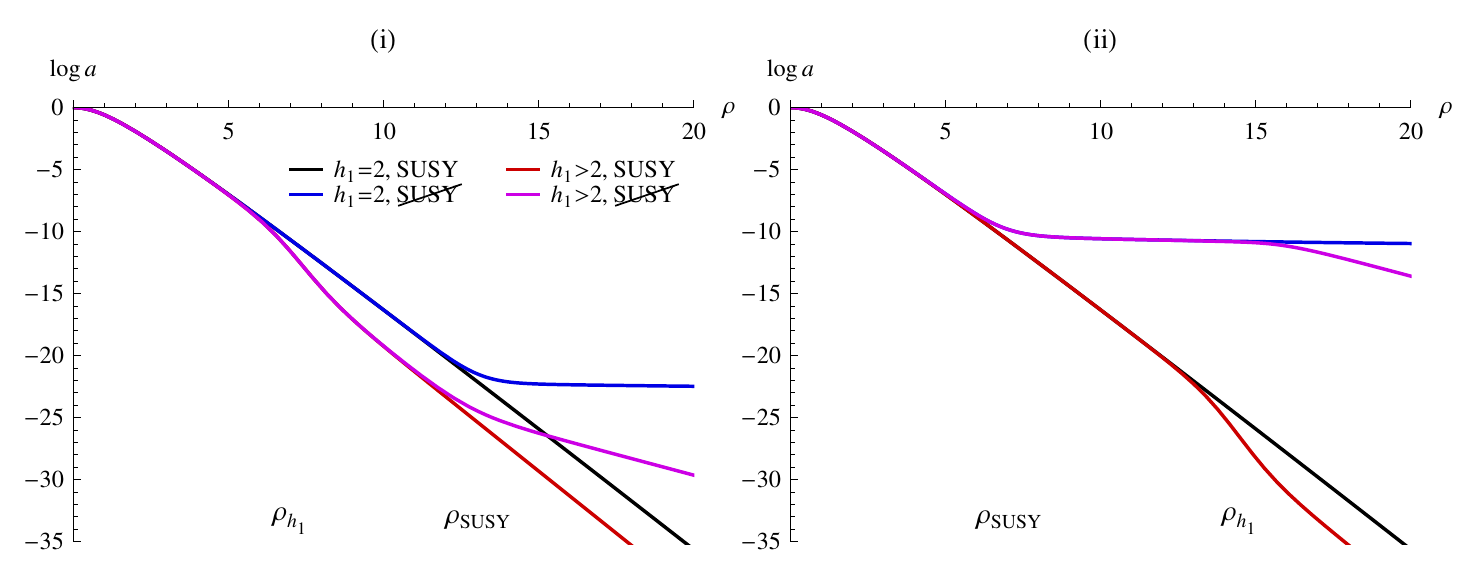}
	\caption{
Plots	of $\log a$ against $\r$ comparing the solutions obtained for each combination of $h_1=2\Nc$, $h_1=2\Nc+\Delta h_1$, $v_2=-2/3$ and $v_2=-2/3+\Delta v_2$.\newline	
(i)		$\Delta h_1 = 10^{-5}$, $\Delta v_2 = 10^{-9}$, with $\Nc=1$. Here $\rMN<\rSUSY$.  In the IR we see the CVMN-like behaviour, with $a\approx 2\r/\sinh2\r$. At $\rMN$ the solutions with $h_1>2$ deviate from this, but after the transition the gradient is unchanged as we still have $a\sim e^{-2\r}$. Then at $\rSUSY$ the non-SUSY solutions switch to the slower decaying behaviour.\newline
(ii)	$\Delta h_1 = 10^{-11}$, $\Delta v_2 = 10^{-4}$, again with $\Nc=1$. Here $\rSUSY<\rMN$. The IR still shows the CVMN-like behaviour. At $\rSUSY$ the non-SUSY solutions switch to the GTV-like behaviour, with $a\sim\r^{-1/2}$. Then at $\rMN$ the solutions with $h_1\ne2$ show a transition. In the SUSY case the gradient is the same after the transition, but in the non-SUSY solution the gradient increases. This corresponds to the transition between $a\sim\r^{-1/2}$ and $a\sim e^{-2\r/3}$.
}	\label{plotScales}
\end{figure}

\begin{figure}[htbp]
	\centering
		\includegraphics{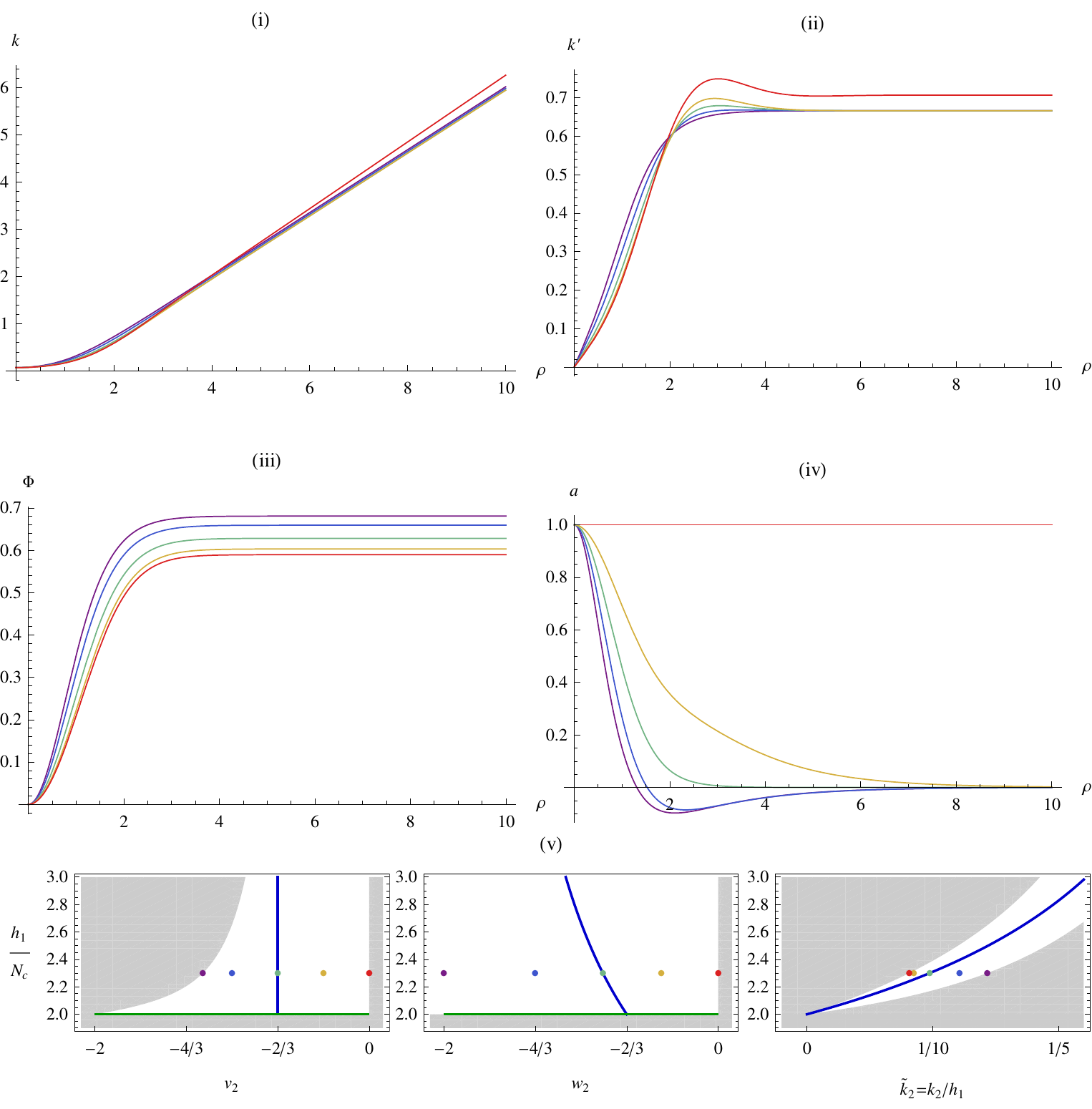}
	\caption{Plots of some of the metric functions for different $v_2$, having set $h_1=2.3$ with $\Nc=1$ and $\phi_0=0$. In (i) we plot $k$, showing that the SUSY-breaking parameter $v_2$ has little effect on the qualitative behaviour except in the case $v_2=0$. This transition between the generic $k\sim2\r/3$ and $k\sim\r/\sqrt{2}$ is shown clearly in (ii), in which we plot the derivative. The UV behaviour, and effect of $v_2$, is very similar in $g$ and $h$. In (iii) we plot the dilaton, showing that $\Phi_\infty$ is a function of $v_2$ for constant $h_1$, and (iv) shows the effect on $a$. The values of $v_2$ used, and the corresponding values of $w_2$ and $k_2$, are shown as coloured points on the solution space diagrams in (v). The colours correspond to those of the curves in (i)--(iv). As in figure \ref{plotParamSpaceSimple}, in (v) the blue curves are the SUSY solutions and the green lines are the GTV solutions. The shaded areas are the regions which are excluded according to the discussion of sections \ref{v2MaxLimit}--\ref{Z2sym}. As we will explain in section \ref{Z2sym}, we can restrict our attention to $w_2>-2$ without loss of generality.}
	\label{plotnonSUSYgeneric}
\end{figure}

In the case of a generic solution, with $h_1>2\Nc$ and $v_2\ne-2/3$, we find that these features survive and both scales are present. The sequence then depends on the ordering of the two scales. If $\rMN<\rSUSY$, the sequence is (figure \ref{plotScales} (i)):
\begin{center}\begin{tabular}{r @{\;} c @{\qquad} l @{\quad} l @{\quad} l }
     $\r<\rMN					$		&		:		&		$	k\approx g \sim \text{constant}	$,	&		$	a\approx b \sim e^{-2\rho}	$		&		(SUSY, CVMN-like)	\\*
     $\rMN<\r<\rSUSY	$		&		:		&		$	k\sim g \sim 2\rho/3						$,	&		$	a\sim b \sim e^{-2\rho}			$		&		(SUSY, KS-like)\\*
     $\r>\rSUSY				$		&		:		&		$	k\sim g \sim 2\rho/3						$,	&		$	a\sim b \sim e^{-2\rho/3}		$		&		(non-SUSY, KS-like)
\end{tabular}\end{center}
On the other hand, if $\rSUSY<\rMN$ we have (figure \ref{plotScales} (ii))
\begin{center}\begin{tabular}{r @{\;} c @{\qquad} l @{\quad} l @{\quad} l }
     $\r<\rSUSY								$		&		:		&		$	k\approx g \sim \text{constant}	$,	&		$	a\approx b \sim e^{-2\rho}	$		&		(SUSY, CVMN-like)	\\*
     $\rSUSY<\r<\rMN					$		&		:		&		$	k\approx g \sim \text{constant}	$,	&		$	a\approx b \sim \r^{-1/2}		$		&		(non-SUSY, GTV-like)\\*
     $\r>\rMN									$		&		:		&		$	k\sim g \sim 2\rho/3						$,	&		$	a\sim b \sim e^{-2\rho/3}		$		&		(non-SUSY, KS-like)
\end{tabular}\end{center}
It appears that $\rMN$ is almost independent of $v_2$, and that $\rSUSY$ is almost independent of $h_1$, although this may break down for sufficiently large $h_1$ and $v_2$, depending on the precise definition used for the scales. In fact the presence of the two scales becomes less clear as they move into the IR for large $h_1$ and $v_2$. This reflects the reduced gradient far from the CVMN solution $(h_1=2\Nc,v_2=-2/3)$ in figures \ref{plotRhoMN} and \ref{plotRhoSUSY}. We show the behaviour of the functions for some generic solutions in figure \ref{plotnonSUSYgeneric}. In this case $h_1$ is large enough that $\rMN$ is not visible.
 
\subsection{The boundaries of the parameter space}\label{v2MaxLimit}
A notable feature of the GTV solutions is the restriction to $-2<v_2<0$ for solutions with a regular UV. There is no obvious way to determine whether an equivalent condition holds for $h_1>2\Nc$, or to find the correct generalisation.

However, numerical observations suggest that $w_2(h_1,v_2)$ becomes independent of $h_1$ for for $v_2\to0$, and that in particular there is a family of solutions with $a=b=1$ and $g=k$ even for $h_1>2\Nc$. This corresponds in our IR expansions \eqref{nonsusyExpIR} to setting
\begin{align}
		w_2 = v_2 = 0,		\qquad\qquad   k_2 = \frac{h_1}{3} - \frac{4\Nc^2}{3h_1} = \frac56 k_2^\text{SUSY},
\end{align}
which agrees with the values obtained numerically. Setting $v_2>0$ (so that $a>1$ for small $\r$) appears numerically to result in a divergent UV, and it seems likely that this is indeed the correct generalisation of the boundary. This corresponds to the solid red curves in figure \ref{plotParamSpace}.

Setting $a=b=1$ and $g=k$ in the equations of motion, we find that our UV ansatz \eqref{nonsusyAnsatzUV}, (expansions in powers of $e^{4\r/3}$) is not suitable. However, using equivalent expansions in powers of $e^{\sqrt{2}\r}$ does lead to a solution:
\begin{align}
e^{2h}&= \frac{\Koo}{2} e^{\sqrt{2} \r}
					+ \left( \frac{\Koo\Kto+\Nc^2}{2\Koo} + \frac{\Nc^2}{2\sqrt{2} \Koo}\r \right)e^{-\sqrt{2}\r}
					+ O\bigl( e^{-3\sqrt{2}\r} \bigr)			\nn\\
e^{2k}&= \Koo e^{\sqrt{2}\r}
					+ \left( \Kto + \frac{\Nc^2}{\sqrt{2}\Koo}\r \right)e^{-\sqrt{2}\r}
					+ O\bigl( e^{-3\sqrt{2}\r} \bigr)			\nn\\
e^{4\Phi-4\Phi_\infty}
			&=	1	-	\frac{1}{\Koo^2}\left(	4\Koo\Kto+\Nc^2 + 2\sqrt{2}\Nc^2 \r		\right) e^{-2\sqrt{2}\r}
						+	O\bigl( e^{-4\sqrt{2}\r} \bigr)																																\label{maxv2uv}
\end{align}
It is important to emphasise that although the form of \eqref{maxv2uv} is simply the original ansatz \eqref{nonsusyAnsatzUV} with the replacement $4\r/3 \to \sqrt{2}\r$, we cannot obtain these expansions from the generic UV \eqref{nonsusyExpUV} simply by a change of coordinates. For example, here we have $e^{2k}=e^{2g}$, whereas in \eqref{nonsusyExpUV} we have $e^{2k}\sim 2e^{2g}/3$ for large $\r$. This is why we have not attempted to match the parameters in \eqref{maxv2uv} to the usual set $\{c_+, c_-, \dots\}$. Instead, we denote the two free parameters by $\Koo$ and $\Kto$, the leading parameter (roughly corresponding to $c_+$) being $\Koo$. Note that, as we have set $v_2=0$, the two parameters $\Koo$ and $\Kto$ cannot be independent once we match to the IR. This is analogous to the SUSY solutions, in which there are two UV parameters $c_+$ and $c_-$, which are related by the requirement to match to the (one-parameter) IR solutions.

In section \ref{GTVlimit} we noted that the `twist' which mixes the $S^2$ and the $S^3$ could be removed by a change of coordinates when $a=b=1$. As we still have $g=k$ here, the same coordinate transformation still works, leading to a simplified system. With $C_4$ and $F_5$ unchanged from \eqref{fieldsRot}, we now find
\begin{align}
	ds_E^2&= e^{\Phi/2}\left[ \hh^{-1/2} dx_{1,3}^2		+		\hh^{1/2}	\left(	e^{2k} d\r^2 + e^{2h} d\Omega_2 + \frac{e^{2k}}{4} d\Omega_3
																																\right)			\right],			\nn\\
	F_3		&= -\frac{\Nc}{4} \wt_1 \wedge \wt_2 \wedge \wt_3,																\nn\\
	H_3		&= 2\Nc e^{2h-2k+2\Phi-\Phi_\infty} \sin\th\ d\r \wedge d\th \wedge d\vph.	
\end{align}

Unfortunately the boundary for $v_2<-2/3$, corresponding to $v_2=-2$ in the GTV solutions, seems to be much less accessible numerically, in part due to the presence of changes of the sign of $a$ and $b$.	However, in the next section we will shed some light on this matter.

\subsection{A $\mathbb{Z}_2$ symmetry}\label{Z2sym}
The system we describe in section \ref{SUSYoverview}, which applies to all the solutions we consider, exhibits a $\mathbb{Z}_2$ symmetry $\I$ which exchanges the two 2-spheres of the conifold and changes the sign of the 3-forms $F_3$ and $H_3$ in \eqref{fieldsRot}. To see this, we make use of the fact that all the systems we consider can be described by the Papadopoulos-Tseytlin ansatz \cite{Papadopoulos:2000gj}. This can be written in the form 
\begin{align}
	ds_E^2 &= e^{\Phi/2} (  \hat{h}^{-1/2}dx_{1,3}^2 + \hat{h}^{1/2}ds_6^2 )								,		\nn\\
	ds_6^2 &=	\frac{2}{3} e^{-8p+3q} ( 4d\r^2 + g_5^2	)
						+ e^{2p+3q} 	\biggl\{	\cosh y\ \Bigl[	e^z		(\w_1^2   + \w_2^2)												
																									+ e^{-z}( \wt_1^2 + \wt_2^2) \Bigr]					\nn\\
				 & \hspace{0.5\textwidth}	-2\sinh y\ (  \w_1 \wt_1 + \w_2 \wt_2)
													\biggr\}																												,		\label{DKMetric}
\end{align}
where the angular forms $g_5$ and $\w_i$ are given by
\begin{align}
	\w_1 = d\th,		\qquad  \w_2 = -\sin\th d\vph,		\qquad g_5 = \wt_3 + \cos\th d\vph.
\end{align}
We use here the notation of \cite{Dymarsky:2011ve}, in anticipation of making contact with their results in section~\ref{DKlimit}.\footnote{
We adapt the notation slightly to avoid confusion with the vielbeins \eqref{vielbeinRot}. The relationship with \cite{Dymarsky:2011ve} is $\w_i^\text{here}  = e_i^\text{\cite{Dymarsky:2011ve}}$ and $\wt_i^\text{here} = \epsilon_i^\text{\cite{Dymarsky:2011ve}}$. We also have $\r_\text{here} = \tau_\text{\cite{Dymarsky:2011ve}} /2$.
}
By comparing the metrics in the two cases we can write an explicit relation between our original functions and those used in \eqref{DKMetric}:
\begin{align}
	e^{10p}		&=	\frac43 e^{g+h-2k}																							,	&
	e^{15q}		&=	\frac38 e^{4g+4h+2k}																						,	\nn\\
	e^y				&=	\frac12 e^{-h} \left( \sqrt{4e^{2h}+e^{2g}a^2} - a e^g \right)	,	&
	e^z				&=	e^{-g} \sqrt{4e^{2h}+e^{2g}a^2}																	.	\label{DKtoOurs}
\end{align}

It is then possible to show that the metric and fields are unchanged (up to a change of sign) if we exchange $(\th,\vph)\leftrightarrow(\tht,\pht)$ and relabel $z \leftrightarrow -z$. In the KS solution \cite{Klebanov:2000hb} which we obtain by taking the limit $h_1,c_+\to\infty$ in the SUSY solutions (section \ref{SUSYsection}), $z=0$ and the transformation $\I$ reduces to a simple change of coordinates. This is the $N_\text{f}=0$ version of the Seiberg duality discussed in~\cite{Casero:2006pt,Casero:2007jj,HoyosBadajoz:2008fw}.

We now consider the effect of $\I$ on a generic globally regular solution of the sort we have discussed, for which $z\ne0$. Inverting \eqref{DKtoOurs} we find
\begin{align}
	e^{2g}	&=				 2^{ 6/5}				e^{ 2p+3q-z}\cosh y				,	&
	e^{2h}	&=				 2^{-4/5}				e^{ 2p+3q+z}\sech y				,	\nn\\
	e^{2k}	&=	\frac{ 2^{11/5} }{3}	e^{-8p+3q}								,	&
	a				&=	e^z \tanh y																			.	\label{oursToDK}
\end{align}
It is then clear that the effect of taking $z\to-z$ can be written as
\begin{align}
	e^{2g}\to e^{2g+2z}		,\qquad\qquad
	e^{2h}\to e^{2h-2z}		,\qquad\qquad
	a			\to e^{-2z}a		.											\label{Z2ours}
\end{align}
Referring to our expansions \eqrefc{nonsusyExpIR}{nonsusyExpUV}, we find
\begin{align}
	e^z = 	\begin{cases}				1	+	(2+w_2)\r^2 + O(\r^4)																														&		\text{for } \r\to0		\\
						\displaystyle				1	+	\frac{1}{c_+} \left( 4H_{11}\r + Q_o + \frac32 c_+ W_{20}^2 \right)e^{-4\r/3}
																+ O\bigl( e^{-4\r/3} \bigr)																												&		\text{for } \r\to\infty .
					\end{cases}		\label{ezExp}
\end{align}
This means that the transformation \eqref{Z2ours} has only subleading effects on $g$, $h$ and $a$, and in particular the transformed functions are still compatible with the form of our expansions \eqrefc{nonsusyAnsatzIR}{nonsusyAnsatzUV}. More specifically, we can see from \eqref{ezExp} that for $z\to-z$ we need to take 
\begin{align}
	w_2 \to -4-w_2,			\label{Z2ir}
\end{align}
corresponding to a reflection in the line $w_2=-2$. As \eqref{susyValsIR} implies that $w_2^\text{SUSY}\to-2$ for $h_1\to\infty$, this is compatible with the fact that the KS solution has $z=0$. Note that because $k$ and $b$ do not appear in \eqref{Z2ours} we can conclude that $k_2$ and $v_2$ are unchanged under $\I$. 

\begin{figure}[htb]
	\centering
		\includegraphics{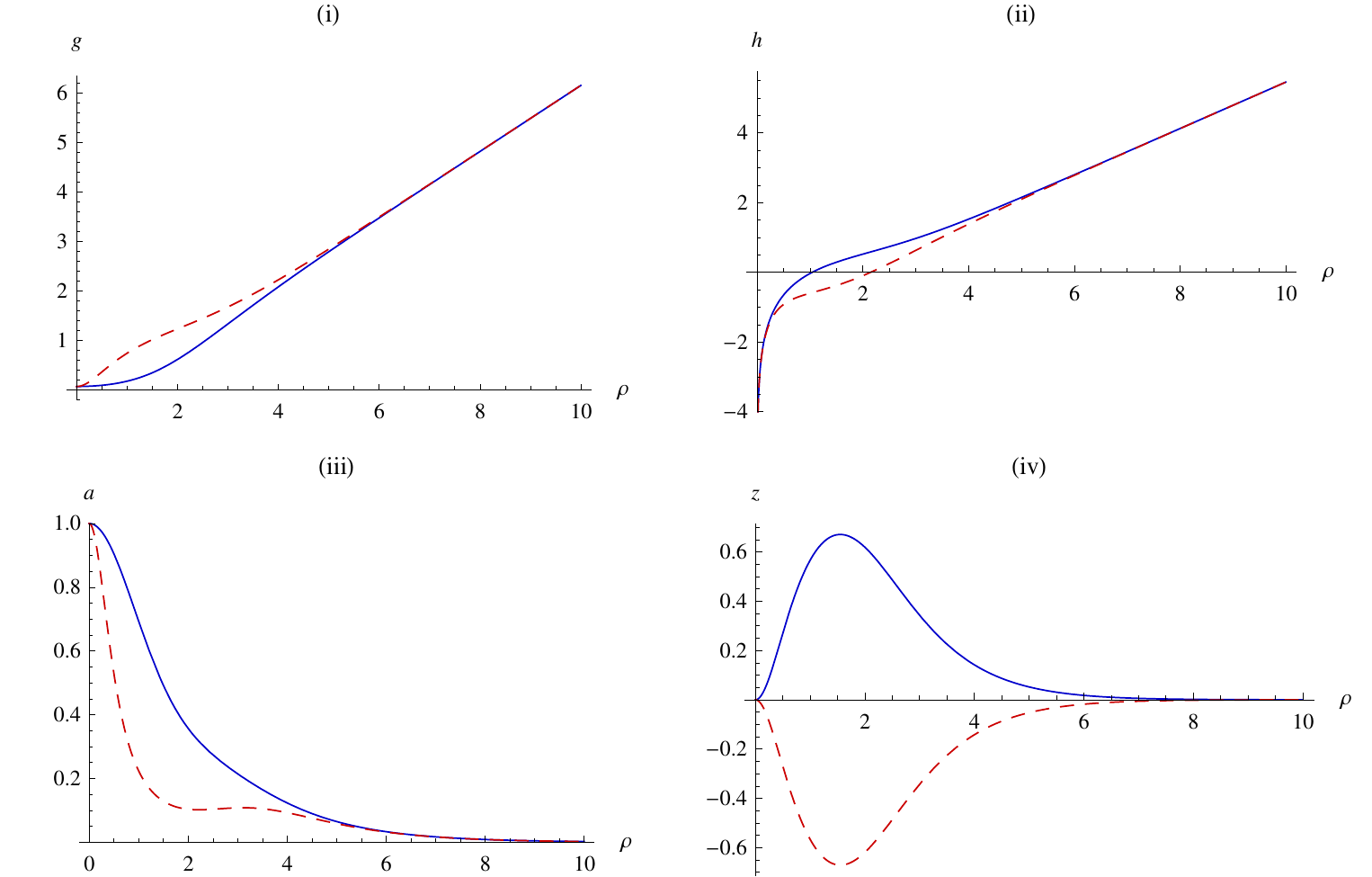}
	\caption{Comparison of the solutions before and after the transformation \eqref{Z2ours}. The blue solid curves correspond to the original description with $w_2>-2$, and the red dashed curves to the `reflected' solutions with $w_2<-2$. In (i)--(iii) we plot the three functions $\{g,h,a\}$ which are affected by the transformation, and in (iv) we show $z$, as defined in \eqref{DKtoOurs}, for which the transformation is simply a change of sign. These solutions have $h_1=2.3$ (with $\Nc=1$) and $v_2=-1/3$, resulting in $k_2\approx0.195$ and $w_2\approx -2 \pm 1.58$ (corresponding to the yellow plots in figure \ref{plotnonSUSYgeneric}).
}
	\label{plotZ2reflection}
\end{figure}

This gives us a simple procedure whereby for each solution discussed in \ref{h1v2dep}, specified by values of $(h_1,v_2)$, we can obtain different solution, with a different value of $w_2$. Because of the way that $\I$ acts only on the subleading terms in the UV expansion we can be sure that the `reflected' solution will also be globally regular and compatible with our ansatz \eqref{nonsusyAnsatzUV}. This can be seen numerically. For a given pair of values of $(h_1,v_2)$, the requirement of UV regularity gives us values of $w_2$ and $k_2$ as described in section \ref{tuningSection}. If we then take $w_2\to-4-w_2$ we immediately find another solution with the correct UV behaviour, without having to adjust $v_2$ or $k_2$. The two types of solutions are compared in figure \ref{plotZ2reflection}.

We should emphasise, however, that although this results in a distinct solution to the equations of motion \eqrefs{eq:eomg}{eq:eomb}, it does not actually correspond to a different background --- $\I$ is simply a relabeling, which is obscured by our choice of basis for the functions.

By demanding that $z\to-z$ while the other functions are unchanged, we can also write down the effect on the UV parameters equivalent to \eqref{Z2ir}:
\begin{align}
 \frac{c_-}{c_+^3}	&\to	\frac{c_-}{c_+^3}		-		32W_{20}
																										\left( 2\frac{Q_o}{c_+}	+ 3     W_{20}^2 													\right)
																										\left( 2e^{2\r_o}				- \frac{Q_o}{c_+} W_{20}	- \frac32 W_{20}^3	\right)		,		\nn\\
   \frac{Q_o}{c_+}	&\to	-\frac{Q_o}{c_+} - 3 W_{20}^2																																					,		\qquad\quad
				 e^{2\r_o}	\to		e^{2\r_o} - \frac{Q_o}{c_+}W_{20} - \frac32 W_{20}^3																									,		\qquad\quad
\frac{H_{11}}{c_+}	&\to	-\frac{H_{11}}{c_+}																																										,		
\label{Z2uv}\end{align}
while keeping the remaining parameters $\{c_+,\Phi_\infty,W_{20},\Phi_{30},V_{40}\}$ fixed. We retain the factors of $1/c_+$ in anticipation of taking the limit $c_+\to\infty$.

For solutions with $w_2=-2$, the IR expansion \eqref{ezExp} appears to vanish at all orders. We would therefore expect that these solutions have $z=0$ for all $\r$, meaning that as in the KS solution $\I$ is a symmetry of the geometry. Our numerical calculations support this assumption --- for these solutions we find that $z$ is indeed essentially zero everywhere (we find $z \lesssim 10^{-14}$ for all $\r\lesssim30$).

\begin{figure}[htbp]
	\centering
		\includegraphics{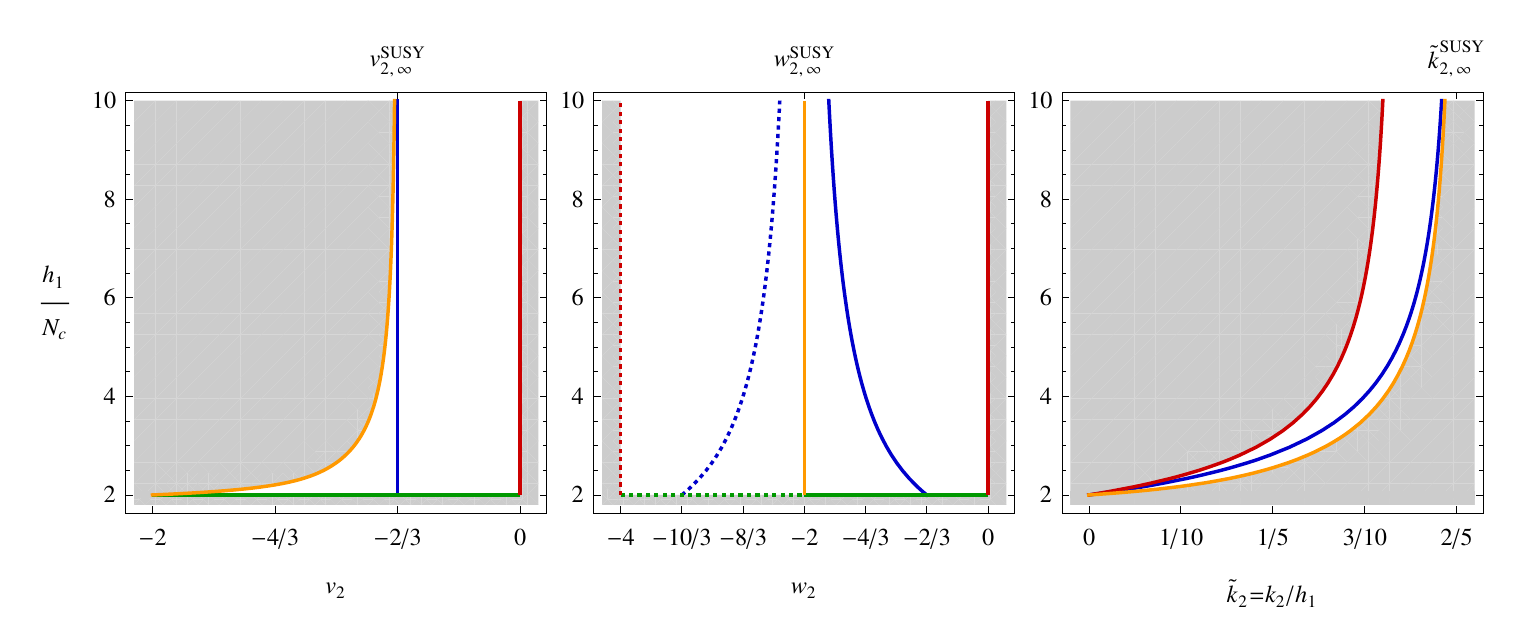}
	\caption{The space of solutions, as in figure \ref{plotParamSpaceSimple}. Again, the blue and green curves are the SUSY and GTV solutions respectively.	The red curves correspond to the case $a=b=1$ discussed in section \ref{v2MaxLimit}, while the orange curves correspond to the solutions which are invariant under $\mathcal{I}$, and so have a $\mathbb{Z}_2$ symmetry of the geometry (section \ref{Z2sym}). The dotted curves are the equivalents with $w_2\to-4-w_2$. 		
	Under the assumption that these two cases constitute the correct generalisation of the requirement $-2\le v_2 \le 0$ in the GTV solutions, the gray shaded areas show the regions where no regular solutions exist. 
	}
	\label{plotParamSpace}
\end{figure}

This family of solutions consists of a line in the $(h_1,w_2)$ plane (see figure \ref{plotParamSpace}), and it would be interesting to determine the corresponding curves in the $(h_1,v_2)$ and $(h_1,k_2)$ planes. We have not been able to determine exact expressions for these functions, but for large $h_1$ we find numerically that	$\Delta v_2(h_1) = v_2(h_1)+2/3 \sim -1/h_1^2$, and 
\begin{align}
	\Delta k_2(h_1) = k_2(h_1) - k_2^\text{SUSY}(h_1) = \frac{16}{45 h_1}   -\epsilon(h_1) ,
\end{align}
where the higher-order corrections $\epsilon>0$ to this last expression are extremely suppressed. For example, with $h_1\approx 10^3$ we find that using $\Delta k_2 = 16/45h_1$ gives the correct value up to around eleven significant digits. In figure \ref{plotParamSpace} the curves $v_2(w_2=-2)$ and $k_2(w_2=-2)$ were obtained from expansions in powers of $1/h_1$ fitted to eight solutions determined numerically.

If these solutions are indeed symmetric under $\I$ for all $\r$ then we can write down a relationship between some of the UV parameters, analogous to the requirement that $w_2=-2$. Specifically, referring to \eqref{Z2uv}, we find\footnote{Of course, we still have the usual undetermined relationships between the UV parameters, so that we are left with only one degree of freedom corresponding to the position on the line $w_2=-2$.}
\begin{align}
	W_{20}^2 = - \frac{2Q_o}{3c_+},		\qquad\qquad  \frac{H_{11}}{c_+} = 0.				\label{Z2conditionUV}
\end{align}
As expected, this is satisfied by the SUSY values \eqref{recoverSUSYUV} in the limit $c_+\to\infty$, corresponding to the KS solution.

In section \ref{v2MaxLimit} we considered the generalisation to $h_1>2\Nc$ of the upper bound $v_2=0$ in the GTV solutions (section \ref{GTVlimit}).
It is suggestive that the line of solutions in which the geometry possesses a $\mathbb{Z}_2$ symmetry passes through the lower bound, $v_2=-2$. In the light of the discussion in this section, we should reinterpret this boundary in the GTV solutions. If we parametrise the solutions by $w_2$, we see that there is no lower bound on $w_2$, but $v_2(w_2)$ has a minimum at $w_2=-2$. This description was not possible in the context of \cite{Gubser2001}, in which all solutions had $a=b$ (so that $v_2=w_2$).

Interpreting the boundary as a minimum of $v_2(w_2)$ would imply that the line $w_2=-2$ is the right generalisation to $h_1>2\Nc$. This is supported by our numerical analysis. It appears not to be possible to tune to a regular UV for values of $v_2$ smaller than that which gives $w_2=-2$.

Of course, we must be cautious here --- our inability to find a solution with $w_2<-2$ could simply be the result of a significant discontinuity in the values of the other parameters across the line $w_2=-2$. 

\subsection{The limit $h_1,c_+ \to \infty$}\label{DKlimit}
Having discussed a non-SUSY generalisation of the baryonic branch, it is natural to consider the generalisation of the Klebanov-Strassler solution \cite{Klebanov:2000hb} itself, which in the SUSY case occurs in the limit $h_1\sim c_+ \to \infty$. In terms of the functions $\{p, q, y, z\}$ which we introduced in section \ref{Z2sym}, the SUSY KS solution has a simple exact description: with $\Phi=\text{constant}$ and $z=0$, we have
\begin{align}
	e^{10p}		=			K^{3} \sinh 2\r														,   \qquad
	e^{15q}		=			\frac{3^{5/4}}{2^{15/2}}K^{2} \sinh^4 2\r	,   \qquad
	e^y				=			\tanh\r																		,						\label{KSexact}
\end{align}
where we have defined
\begin{align}
		K			\equiv	\frac{(\sinh 4\r - 4\r )^{1/3}}{2^{1/3}\sinh 2\r}.
\end{align}
The remaining function $b=2\r/\sinh2\r$ is the same as in the whole SUSY baryonic branch. As we have seen in section \ref{Z2sym}, the fact that $z=0$ implies that the geometry itself possesses a $\mathbb{Z}_2$ symmetry.

Of course, in order for the concept of a non-SUSY generalisation to be meaningful, we have to choose which characteristics of the SUSY KS solution we want to keep in the non-SUSY solution. One natural possibility would be to require that the geometry retains the $\mathbb{Z}_2$ symmetry, in which case the we obtain the family of solutions with $w_2=-2$ which we discussed in section \ref{Z2sym}.

However in \cite{Dymarsky:2011ve}, Dymarsky and Kuperstein (DK) followed a different approach. They noted that the KS background has several simplifying features which are retained in the linear deformations studied in \cite{Gubser:2004qj,Gubser:2004tf}, but not in the generic baryonic branch:
\begin{enumerate}[(i)]
	\item A constant dilaton, $e^\Phi = g_s$
	\item An imaginary self-dual 3-form flux\footnote{Here $\ast_6$ is the six-dimensional Hodge dual}, $iG_3 = \ast_6 G_3$, where $G_3 \equiv F_3+ \frac{i}{g_s}H_3$
	\item An RR 4-form satisfying $C_4 = H^{-1} \mathrm{Vol}_{1,3}$, where $ds^2 = H^{-1/2}dx_{1,3}^2+ H^{1/2} ds_6^2$
	\item A Ricci-flat 6d unwarped metric 
\end{enumerate}
As noted in \cite{Dymarsky:2011ve}, these are particularly convenient because they mean that the fluxes completely decouple from the equations which determine the metric. It should be noted that in our solutions (ii) and (iii) are satisfied automatically once (i) is imposed. 

By imposing that these properties are retained, DK found a one-dimensional family of solutions which break both SUSY and the $\mathbb{Z}_2$ symmetry of the geometry (although the full symmetry including the exchange $z\leftrightarrow-z$ is of course retained). It seems natural to assume that this corresponds to a line of solutions in the two-dimensional solution space described above.

To see that this is indeed the case, we first need to identify the appropriate limit. Referring to our generic IR expansions \eqref{nonsusyExpIR}, we see that we obtain a constant dilaton in the limit $h_1\to\infty$, as in the SUSY case. This means that conditions (i)--(iii) are satisfied. It is also possible to check that this results in the IR expansion for the 6d Ricci scalar vanishing, as required by condition (iv).

We now look to relate our three SUSY-breaking parameters $\{w_{2},k_{2},v_{2}\}$ to the parameters $\{\zeta_{1},\zeta_{2},\zeta_{3}\}$ used in \cite{Dymarsky:2011ve}. 
Looking then at the IR expansion for $z$, we find by substituting our IR expansions \eqref{nonsusyExpIR} into \eqref{DKtoOurs}
\begin{align}
z = (2+w_2) \r^2 + O(\r^4),			\label{DKir:z}
\end{align}
meaning we can compare with the expression given in \cite{Dymarsky:2011ve} and conclude that
\begin{align}
w_{2}=4\zeta_{1}-2.
\end{align}
To gain the relation for $k_{2}$ we look at the expansion for $e^{y}$ and upon taking the limit $h_{1}\to\infty$ we find
\begin{align}
	e^y = \r - \left(\frac23 + 4\zeta_1^2 \right)\r^3 + O(\r^5).
\end{align}
This does not have enough freedom in the $\rho^{3}$ term when compared to \cite{Dymarsky:2011ve}. To fix this, it is possible to take $k_{2}\to\infty$ while keeping fixed $\tilde{k}_{2}\equiv k_{2}/h_{1}$. This then gives
\begin{align}
	e^y = \r - \left(\frac23 + 4\zeta_1^2 - \frac56 \tilde{k}_2 \right )\r^3 + O(\r^5),
\end{align}
which we can match to the result of \cite{Dymarsky:2011ve} by setting
\begin{align}
	\tilde{k}_2 \equiv \frac{k_2}{h_1} = \frac{2}{65} (13 - 90 \zeta_2).
\end{align}
We finally need to determine the relationship between $v_{2}$ and $\zeta_{3}$. This can be achieved by comparing our expansion for $b$ with that for 
$F=(1-b)/2$ in \cite{Dymarsky:2011ve}, from which we obtain
\begin{align}
	v_2 = -\frac23 (\zeta_3 + 1).
\end{align}
In summary, in the limit $h_1\to\infty$ we find the following relationships between our three SUSY-breaking IR parameters and those used in \cite{Dymarsky:2011ve}:
\begin{align}
	w_2 = 4\zeta_1 - 2,																											\qquad
	\tilde{k}_2 \equiv \frac{k_2}{h_1} = \frac{2}{65} (13 - 90 \zeta_2),		\qquad
	v_2 = -\frac23 (\zeta_3 + 1).
\end{align}

Of course, setting the $\zeta_i$ to zero we recover (the large-$h_1$ limit of) the SUSY values \eqref{susyValsIR}. In fact, defining for example $\Delta w_2(h_1) = w_2 - w_2^\text{SUSY}(h_1)$, we obtain
\begin{align}
	\zeta_1	=	 \frac14 \Delta w_2									,		\qquad
	\zeta_2 = -\frac{13}{36} \Delta\tilde{k}_2		,		\qquad
	\zeta_3 = -\frac32 \Delta v_2.
\end{align}

In the UV we are less sure how to find similar relationships between parameters. It is clear from the numerical analysis that the relevant limit is still $c_+\to\infty$ (even if the precise relation \eqref{cph1} may no longer hold in the non-SUSY case), and we know we will need $\Phi_\infty \to\phi_0$ in order to get a constant dilaton. However, it is not obvious how the other parameters in \eqref{UVparam} behave in this limit. One possibility is suggested by the fact that in the case of the IR parameters we could have guessed the correct behaviour ($v_2\sim w_2\sim\text{constant}$, $k_2\sim h_1$) from the $h_1$-dependence of the SUSY values \eqref{susyValsIR} in the limit. Using the same approach in the UV would imply that we should consider all the remaining parameters fixed except for $\Phi_{30} \sim 1/c_+^2$.  

Looking at the UV expansions for the 6d Ricci scalar and the dilaton we find that in fact the limit $c_+\to\infty$ is itself sufficient for Ricci-flatness, and taking both $c_+\to\infty$ and $\Phi_{30}\to0$ gives a constant dilaton. This can be seen for the SUSY baryonic branch in figure \ref{plotSUSY} (iv); the non-SUSY solutions show qualitatively the same behaviour.

Unlike in the case $h_1\to2\Nc$, our numerical approach does not allow us to take the limit $h_1\to\infty$ explicitly. However, we can probe sufficiently large values of $h_{1}$ to yield solutions which appear to have many of the characteristics we expect from the true limit. For example, we do not have to take $h_1$ very large before the dilaton is very close to constant. Notice in figure \ref{plotSUSY} (iv) the curve for $h_1=12$ appears to lie on the axis.


\section{Remarks on the dual field theory}\label{fieldTheory}
Here we shall discuss a little about the dual field theories to the gravity backgrounds we have presented.  Much of the following is similar to that of \cite{Bennett:2011va}, although we have not restricted ourselves to small deformations, as was discussed in that paper.

We will only consider the solutions with $h_1>2\Nc$. In this case the geometry is `almost' asymptotically $AdS_5$. More precisely, for large $\r$ we can write the metric in the form
\begin{align}
	ds^2 &\sim \frac{u^2}{H(u)^{1/2}} dx_{1,3}^2 + \frac{H(u)^{1/2}}{u^2} du^2 + ds_5^2, 	\nn\\
	H(u) &\sim \log u + \text{constant} + O(u^{-2}),																							\label{AdSmetric}
\end{align}
where we have defined a suitable radial coordinate $u$ (increasing with $\r$). For the generic solutions satisfying the ansatz \eqref{nonsusyAnsatzUV} (including the SUSY solutions), the definition which results in \eqref{AdSmetric} is $u = e^{2\r/3}$. For the solutions with $v_2=0$ discussed in section \ref{v2MaxLimit} we instead need $u=e^{\r/\sqrt{2}}$. The term of order $\log u$ in the correction $H(u)$ results from the sub-leading behaviour of the dilaton \eqrefc{nonsusyExpUV}{maxv2uv}.

There are three different field combinations which are invariant under the rotation which are of interest \cite{Elander:2011mh}. The first is the dilaton $\Phi$, and the others are defined as
\begin{align}
	M_{1}= e^{2z}-1 = a^{2} + 4 e^{2h-2g}-1									,\qquad
	M_{2}= e^{2h+2g-4k}.																								\label{M1M2def}
\end{align}

In the case of the generic solutions described by \eqref{nonsusyExpUV}, these functions have UV expansions
\begin{align}
	e^{\Phi-\Phi_\infty}&= 1-\left(\frac{3\Nc^2 }{2c_+^2}\r-e^{-4\Phi_\infty} \frac{\Phi_{30}}{4}\right) e^{-\frac83\r}+O\bigl(     e^{-4\r  } \bigr),			\nn\\
	M_{1}&= \left(8H_{11}\r + 3 c_{+} W_{20}^{2} + 2 Q_{0}\right)\frac{e^{-4\r/3}}{c_{+}}+O(e^{-8\r/3})	, \nn\\
	M_{2}&= \frac{9}{16} - \frac{27}{16} W_{20}^{2} e^{-4\r/3}+O(e^{-8\r/3}).																				\label{5dfields}
\end{align}

By looking at the asymptotic behaviour of fields (and combinations of them) it is possible to think of our constants in terms of the operators which are deforming a fixed point. We may do this as it is understood that a generic field $\mathcal{M}\sim u^{-\Delta}$ as $u\to\infty$ behaves in the following manner. If $\D > 0$ (or $\D=0$) it is either an indication of a relevant (or marginal) operator in the Lagrangian or the VEV for an operator of dimension $\D$. If instead, $\D < 0$, then it indicates the insertion of an irrelevant operator of dimension $(4-\D)$ in the Lagrangian.

Using this analysis it can be seen, from the UV expansion above, that the dilaton falls into the marginal operator category as it has scaling dimension $\D = 4$ (this can be associated with a certain combination of gauge couplings discussed in \cite{Bennett:2011va}).

We can further use this analysis on the expansion of the function $b(\r)$ presented here for convenience
\begin{align}
b= \frac{9W_{20}}{4} e^{-\frac23\r}+ \left[\frac{10W_{20}^3}{3}	\r^2+ \left(	4e^{2\r_o} -	\frac{Q_o W_{20}}{c_+}
-	\frac{23W_{20}^3}{6}	\right)\r+	V_{40}	\right]e^{-2\r}+		O\bigl( e^{-\frac{10}{3}\r} \bigr)  	.
\end{align}

Here we can see that $W_{20}$, which we could consider to be our `SUSY-breaking constant', corresponds to an operator of dimension three being inserted in the Lagrangian.  We anticipated in section \ref{generalNonSUSY} that we can associate this operator with the mass of the gaugino, as in \cite{Bennett:2011va}. Following the SUSY case we also associate $e^{2\r_{0}}$, which appears at next-to-leading order in $M_1$, with the VEV of the gaugino. From this we can write schematically
\begin{align}
W_{20} \to m\l \l, \qquad e^{2\r_{0}}\to \<\l \l\> \sim \L^{3}_\text{YM}.
\end{align}
It should be noted that this association is not exact once we have broken SUSY --- the SUSY-breaking parameter can generically also deform the gaugino VEV, as indicated by the contributions from $W_{20}$ and $V_{40}$ to $M_1$ in \eqref{5dfields}.

As discussed in appendix \ref{matchingAppendix}, it appears that $W_{20}\to\infty$ as we approach the boundary at $v_2=w_2=0$ (figure \ref{plotUVparam}). This suggests that we can interpret the solution \emph{on} the boundary, with $a=b=1$ for all $\r$ (section \ref{v2MaxLimit}) as corresponding to a field theory in which the gaugino has been given infinite mass. We therefore no longer have soft SUSY breaking --- the theory is non-SUSY all the way into the UV. Presumably, by sending the mass to infinity we effectively remove the gaugino entirely, obtaining a completely non-SUSY theory.

We can now look to the field combination $M_{1}$ and see that it can be thought of as corresponding to the VEV of a dimension two operator $\mathcal{U}$. In the SUSY case we can identify~\cite{Dymarsky:2005xt} 
\begin{align}
\mathcal{U} \sim \tr[AA^{\dag}-B^{\dag}B],	\label{UAABB}
\end{align}
and this operator getting a VEV is the exact thing which allows us to explore the baryonic branch. Notice that in the SUSY case $W_{20}=0$ and the leading term of $M_1$ vanishes for $c_+\to\infty$, when we recover KS. This is also the limit in which the geometry is invariant under the $\mathbb{Z}_2$ symmetry $\I$ which we discussed in section \ref{Z2sym}. In fact, from the point of view of the field theory, the transformation $\I$ can be identified with swapping $A\leftrightarrow B$ \cite{Gubser:2004qj}.

As soon as we move away from the SUSY solutions we can no longer make the identification \eqref{UAABB}. However, it is still instructive to consider the behaviour of the operator $\mathcal{U}$ associated with $M_1$. From \eqref{5dfields} it is clear that we can expect $\mathcal U$ to be changed when we break SUSY while keeping $c_+$ fixed. Indeed, referring to the definition \eqref{M1M2def}, we see that $M_1=0$ when $z=0$. This applies at all $\r$ in all the solutions on the line $w_2=-2$. (As required, we see that the combination of parameters appearing in the UV expansion \eqref{5dfields} vanishes when \eqref{Z2conditionUV} is satisfied.) It is interesting that the presence of the $\mathbb{Z}_2$ symmetry still corresponds to the vanishing of this operator, even in the non-SUSY case. This is perhaps indicative of the extent to which the structure of the SUSY system survives in the generic case.

As we move in the opposite direction from the SUSY solutions, increasing $v_2$ (and $W_{20}$) we find numerically (appendix \ref{matchingAppendix}) that both terms at leading order in $M_1$ diverge. However, in the limit we obtain the solutions described in section~\ref{v2MaxLimit} and the expansions \eqref{5dfields} are no longer valid. Instead, for large $\r$ 
\begin{align}
	M_1 = 2  +  \frac{2\Nc^2}{\Koo^2} e^{-2\sqrt{2}\r} + O\bigl( e^{-4\sqrt{2}\r} \bigr).
\end{align}
This is qualitatively different to the generic case. Firstly, we now have $M_1\to2$ in the UV, as opposed to $M_1\to0$. This indicates that these solutions do not recover the $\mathbb{Z}_2$ symmetry in the UV. Secondly, the next-to-leading term is now of order $u^{-4}$, meaning that we can no longer associate this field with a dimension two operator. 

There is some subtlety here in the fact that unlike in \cite{Bennett:2011va} we have allowed our deformations of the SUSY solutions to become large. It is then not clear that any deductions based on analogy with the SUSY solutions remains valid. In particular, we cannot not necessarily expect to find stable solutions for all values of $W_{20}$. However, the similarities between the SUSY and non-SUSY solutions are interesting. It should be noted that we still find a continuous and smooth deformation of the SUSY solutions between smaller and larger values of the non-SUSY deformations in the IR. We only find a different UV expansion in the limiting cases (or boundaries of our solution space).


\section{Summary and conclusions}

In this paper we study the full two-dimensional space of solutions which can be considered to be the non-SUSY generalisation of the baryonic branch (extending the work of \cite{Bennett:2011va}). We include the solutions compatible with the PT ansatz which have both a regular IR, of the same form as that of the baryonic branch, and are related to the baryonic branch by a continuous change of parameters.

In addition to the SUSY baryonic branch and its limiting cases (Klebanov-Strassler and Chamseddine-Volkov/Maldacena-N\'u\~nez), this solution space also includes two previously studied one-dimensional families of non-SUSY solutions as limits. In the limit which yields in the SUSY case the CVMN solution we obtain the solutions of Gubser, Tsyetlin and Volkov \cite{Gubser2001} (presented here in section \ref{GTVlimit}), while in the limit corresponding to the KS solution itself we obtain those of Dymarsky and Kuperstein \cite{Dymarsky:2011ve} (presented in section \ref{DKlimit}).  The behaviour of generic non-SUSY solutions lying away from these boundaries can be understood as a combination of the effects which are present in the SUSY baryonic branch and the GTV solutions.

Alongside these cases we identify two additional one-dimensional families which are of interest. The first is the boundary of the solution space with $v_2=w_2=0$ corresponding to the positive boundary of the GTV solutions. Here we can no longer argue that SUSY is softly broken (the gaugino mass appears to be infinite), and we find that $a=b=1$ for all $\r$. Notably, this changes the geometry to an explicity non-SUSY case (a cone over $S^{2} \times S^{3}$). We also find an explicit UV expansion for the solutions on this boundary which is different from the generic UV.  The second family lies on the line $w_2=-2$, upon which the geometry possesses a $\mathbb{Z}_2$ symmetry just as in the Klebanov-Strassler solution. This family of solutions corresponds to the other boundary of the GTV solutions. 

Moving away from the boundaries, we have also shown that solutions with $w_2<-2$ are related to those with $w_2>-2$ by a $\mathbb{Z}_2$ symmetry and describe the same physical system, although the solutions themselves appear different. In the two-dimensional solution space much of the SUSY structure survives. In addition to the various quantities calculated in \cite{Bennett:2011va}, which are mostly unaffected by SUSY-breaking at leading order, we find that the presence of a $\mathbb{Z}_2$ symmetry of the geometry is still linked to the vanishing of a dimension-two operator. In the SUSY case this reflects the fact that in the dual field theory the $\mathbb{Z}_2$ transformation corresponds to the ability to interchange the baryons.

It would be interesting to know to what extent this description applies to the non-SUSY case. To address this, it would be necessary to gain a more detailed understanding of the field theory, including calculation of the mass spectrum. Another question which we did not address is the issue of stability. It would be useful to determine if, and how much, the parameter space is restricted by this requirement. Finally, we note that the transition between the generic UV \eqref{nonsusyExpUV} and the boundary case \eqref{maxv2uv} is somewhat unclear. It appears that the solutions first approach the boundary case before switching to the generic behaviour in the UV, the scale at which this occurs presumably being associated with the gaugino mass. However, more detailed study of the solutions in this region would be necessary to understand this completely.

\section*{Acknowledgments}
We would like to thank Carlos N\'u\~nez for extensive discussions. This work was supported by STFC studentships.


\appendix

\makeatletter
\def\@seccntformat#1{\csname Pref@#1\endcsname \csname the#1\endcsname\quad}
\def\Pref@section{Appendix~}
\makeatother

\section{Equations of motion}\label{EOMappendix}
The equations of motion for the full non-SUSY system (section \ref{generalNonSUSY}) can be obtained either from the Einstein, Maxwell,
dilaton and Bianchi equations of the ten-dimensional system, or from a one-dimensional effective Lagrangian $L=T-U$, with
\begin{align}
T  &=		-\frac{1}{128} e^{2 \Phi } \Bigl\{e^{4 g} \left(a'\right)^2+\left(b'\right)^2 N_c^2-8 e^{2 (g+h)} \Bigl[2 g' \left(2 h'+k'+2 \Phi '\right)+\left(g'\right)^2		\nn\\
	&\qquad\quad		+2 h' \left(k'+2 \Phi '\right)+\left(h'\right)^2+2 \Phi ' \left(k'+\Phi '\right)\Bigr]\Bigr\}   , \nn\\ 
U	 &=		\frac{1}{256} e^{-2 (g+h-\Phi )} \Bigl[a^4 e^{4 g} \left(N_c^2+e^{4 k}\right)-4 a^3 b e^{4 g} N_c^2+2 a^2 e^{2 g} \Bigl(2 b^2 e^{2 g} N_c^2	\nn\\
	&\qquad\quad		+e^{2 g} N_c^2+4 e^{2 h} N_c^2-8 e^{2 (g+h+k)}+4 e^{4 g+2 h}-e^{2 g+4 k}+4 e^{2 h+4 k}\Bigr)		\nn\\
	&\qquad\quad		-4 a b e^{2 g} N_c^2 \left(e^{2 g}+4 e^{2 h}\right)+8 b^2 N_c^2 e^{2 (g+h)}+e^{4 g} N_c^2	+16 e^{4 h} N_c^2	\nn\\
	&\qquad\quad		-16 e^{2 (2 g+h+k)}-64 e^{2 (g+2 h+k)}+e^{4 (g+k)}+16 e^{4 (h+k)}\Bigr]	.		
\label{1dL}
\end{align}
In addition to the equations of motion resulting from \eqref{1dL}, there is a Hamiltonian constraint $T+U=0$ resulting from invariance under reparametrisation of the radial coordinate.

The equations of motion themselves, setting $\Nc=1$ for simplicity, are
\begin{align}
	g''   &=  \frac{1}{8} e^{-4 g-2 h} \Bigl[e^{6 g} \left(a'\right)^2-4 a^2 e^{2 g+4 k}-4 a^2 e^{2 g}+4 a^2 e^{6 g}+8 a b e^{2 g}	\nn\\
	&\qquad\quad		-e^{2 g} \left(b'\right)^2-4 b^2 e^{2 g}-16 e^{4 g+2 h} g' h'-16 e^{4 g+2 h} g' \Phi'		\nn\\
	&\qquad\quad		-16 e^{4 g+2 h} \left(g'\right)^2+32 e^{2 g+2 h+2 k}-16 e^{2 h+4 k}-16 e^{2 h}\Bigr]    
\label{eq:eomg}\\
  h''   &=  -\frac{1}{8} e^{-2 g-4 h} \Bigl[\left(a'\right)^2 e^{4 g+2 h}+a^4 e^{2 g+4 k}+a^4 e^{2 g}-4 a^3 b e^{2 g}+4 a^2 b^2 e^{2 g}	\nn\\
	&\qquad\quad		-8 a^2 e^{2 g+2 h+2 k}+4 a^2 e^{4 g+2 h}-2 a^2 e^{2 g+4 k}+2 a^2 e^{2 g}	\nn\\
	&\qquad\quad		+4 a^2 e^{2 h+4 k}+4 a^2 e^{2 h}-4 a b e^{2 g}-8 a b e^{2 h}+e^{2 h} \left(b'\right)^2	\nn\\
	&\qquad\quad		+4 b^2 e^{2 h}+16 e^{2 g+4 h} g' h'+16 e^{2 g+4 h} h' \Phi'+16 e^{2 g+4 h} \left(h'\right)^2	\nn\\
	&\qquad\quad		-8 e^{2 g+2 h+2 k}+e^{2 g+4 k}+e^{2 g}\Bigr]    \\ 
	k''   &=  \frac{1}{8} e^{-4 g-4 h} \Bigl(a^4 e^{4 g+4 k}-a^4 e^{4 g}+4 a^3 b e^{4 g}-4 a^2 b^2 e^{4 g}+8 a^2 e^{2 g+2 h+4 k}	\nn\\
	&\qquad\quad		-8 a^2 e^{2 g+2 h}-8 a^2 e^{6 g+2 h}-2 a^2 e^{4 g+4 k}-2 a^2 e^{4 g}+16 a b e^{2 g+2 h}	\nn\\
	&\qquad\quad		+4 a b e^{4 g}-8 b^2 e^{2 g+2 h}-16 e^{4 g+4 h} g' k'-16 e^{4 g+4 h} h' k'	\nn\\
	&\qquad\quad		-16 e^{4 g+4 h} k' \Phi'+e^{4 g+4 k}-e^{4 g}+16 e^{4 h+4 k}-16 e^{4 h}\Bigr)     \\
	\Phi''&=  \frac{1}{8} e^{-4 g-4 h} \Bigl[a^4 e^{4 g}-4 a^3 b e^{4 g}+4 a^2 b^2 e^{4 g}+8 a^2 e^{2 g+2 h}-16 a b e^{2 g+2 h}	\nn\\
	&\qquad\quad		+2 a^2 e^{4 g}-4 a b e^{4 g}+2 \left(b'\right)^2 e^{2 g+2 h}+8 b^2 e^{2 g+2 h}-16 e^{4 g+4 h} g' \Phi'	\nn\\
	&\qquad\quad		-16 e^{4 g+4 h} h' \Phi'	-16 e^{4 g+4 h} \left(\Phi'\right)^2+e^{4 g}+16 e^{4 h}\Bigr]    \label{eq:eomphi} \\
	a''   &=  e^{-4 g-2 h} \Bigl(-4 a' e^{4 g+2 h} g'-2 a' e^{4 g+2 h} \Phi'+a^3 e^{2 g+4 k}+a^3 e^{2 g}-3 a^2 b e^{2 g}	\nn\\
	&\qquad\quad		+2 a b^2 e^{2 g}-8 a e^{2 g+2 h+2 k}+4 a e^{4 g+2 h}-a e^{2 g+4 k}+a e^{2 g}	\nn\\
	&\qquad\quad		+4 a e^{2 h+4 k}+4 a e^{2 h}-b e^{2 g}-4 b e^{2 h}\Bigr)    \\ 
	b''   &=  -e^{-2 h} \left(a^3 e^{2 g}-2 a^2 b e^{2 g}+a e^{2 g}+4 a e^{2 h}+2 e^{2 h} b' \Phi'-4 b e^{2 h}\right)	\label{eq:eomb}
\end{align}

The case discussed in section \ref{v2MaxLimit}, with $v_2=0$, is far simpler. After setting $a=b=1$ and $g=k$ the equations of motion for the remaining three functions are
\begin{align}
	k''			&=	2 - 2e^{-4k} - 2h' k' 	- 2 (k')^2 	- 2 k' \Phi'		,	\nn\\
	h'' 		&= e^{2k-2h} 	 - 2h' k' 	- 2 (h')^2 	- 2 h' \Phi'			,	\nn\\
	\Phi'' 	&= 2 e^{-4k} 		 - 2h'\Phi' - 2(\Phi')^2- 2 k' \Phi' 		,
\end{align}
and the constraint is
\begin{align}
	e^{-4k} - e^{2k-2h} -3  + 6h'k'  + 4h'\Phi' + 6k'\Phi' + (h')^2 + 3(k')^2+ 2(\Phi')^2		=	0		.
\end{align}

%
\section{Obtaining the UV parameters}\label{matchingAppendix}
To look for the UV behaviour of the solutions in \cite{Bennett:2011va}, a matching procedure was proposed to provide a fit of the UV parameters.  However, we find this process is unreliable when we match at large $\r$. Here we are interested in looking at solutions where the the two scales $\rMN$ and $\rSUSY$ are varied over a large range. In particular, we need to include cases in which one or both have large values. The correct UV behaviour for solutions of this type is only manifest at large $\r$, meaning that the matching procedure used in \cite{Bennett:2011va} is unsuitable.

\begin{figure}[htbp]
	\centering
		\includegraphics{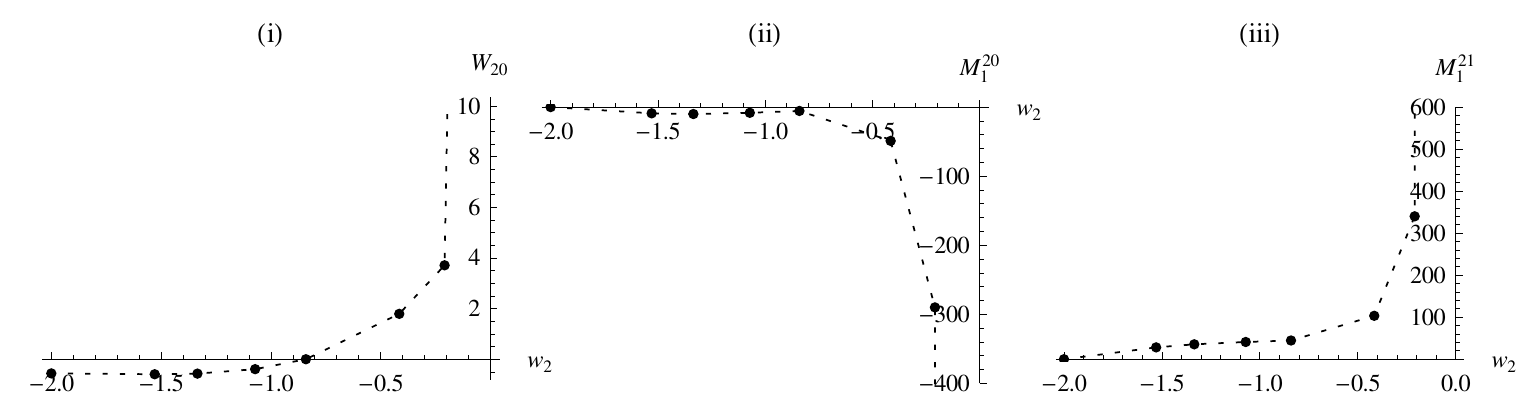}
	\caption{ Plots of the some of the UV parameters, estimated using the method described above for $h_1=2.3$ with $\Nc=1$ and $\phi_0=0$ (this includes the solutions plotted in figure \ref{plotnonSUSYgeneric}). In (i) we plot $W_{20}$, corresponding to the gaugino mass which breaks SUSY. We also include (ii) $M_1^{20}\equiv 3W_{20}^2 + 2Q_o/c_+$ and (iii) $M_1^{21}\equiv 8H_{11}/c_+ $, which contribute to the leading term in $M_1$ discussed in section \ref{fieldTheory}. The dotted lines diverging for $w_2\to0$ indicate the position of the next point, at $w_2\approx 1.3\times10^{-3}$. This has $W_{20}\approx 10^2$, $M_1^{20}\approx -6\times10^5$ and $M_1^{21}\approx 3\times10^5$.}
	\label{plotUVparam}
\end{figure}

Instead of performing this full matching procedure, it is possible to estimate some parameters from the leading behaviour of appropriate combinations of the background functions. For example, we can use the combination 
\begin{align}
-\frac{3}{2} e^{2\r/3}a'(\r) \to W_{20}
\end{align}
to give an approximation of the SUSY breaking parameter in cases where the matching procedure fails. Using this method, we find that it appears that $W_{20}\to\infty$ for $v_2\to0$ (see figure \ref{plotUVparam}). The case $w_2=-2$ has $W_{20}'(w_2)=0$, as would be expected from the invariance of $W_{20}$ under the transformation \eqref{Z2uv}.

Using the same method, the leading coefficients in $M_1$ (see \eqrefc{M1M2def}{5dfields}), can be seen to have similar behaviour. However, both these quantities vanish for $w_2=0$, with probably non-zero derivatives. This reflects the fact that their signs change under the transformation \eqref{Z2uv}.

\renewcommand\bibname{References}
\bibliography{nonSUSYbib}

\end{document}